\documentclass[12pt]{iopart}
\usepackage{graphicx}
\eqnobysec

\newcommand{\beq}{\begin{equation}}
\newcommand{\beqa}{\begin{eqnarray}}
\newcommand{\eeq}{\end{equation}}
\newcommand{\eeqa}{\end{eqnarray}}
\newcommand{\abs}[1]{\vert{#1}\vert}

\renewcommand{\d}{{\rm d}}

\renewcommand{\e}{{\rm e}}
\newcommand{\frad}[2]{\displaystyle{\displaystyle#1\over\displaystyle#2}}

\newcommand{\point}{{\noindent $\bullet$ }}

\renewcommand{\Re}{\mathop{{\rm Re}}}
\renewcommand{\Im}{\mathop{{\rm Im}}}

\begin{document}

\title[] 
{Quantum scattering by a disordered target ~---~ The mean cross section} 

\author{D Boos\'e$^{1,2}$, J Y Fortin$^{2}$ and J M Luck$^{3}$}

\address{$^{1}$ Institut Denis Poisson, Laboratoire de Math\'ematiques et Physique Th\'eorique,\\  
Universit\'e de Tours, CNRS UMR 7013, 37200 Tours Cedex, France}

\address{$^{2}$ Laboratoire de Physique et Chimie Th\'eoriques, Universit\'e de Lorraine,\\
CNRS UMR 7019, 54506 Nancy Cedex, France}

\address{$^{3}$ Universit\'e Paris-Saclay, CNRS \& CEA, Institut de Physique Th\'eorique,
91191~Gif-sur-Yvette, France}

\eads{\mailto{dominique.boose@lmpt.univ-tours.fr},\mailto{jean-yves.fortin@univ-lorraine.fr},
\mailto{jean-marc.luck@ipht.fr}}

\begin{abstract}
\\
We study the variation of the mean cross section with the density of the samples in
the quantum scattering of a particle by a disordered target. The target consists of a
set of pointlike scatterers, each having an equal probability of being anywhere inside
a sphere whose radius may be modified. We first prove that scattering by a pointlike
scatterer is characterized by a single phase shift ${\delta}$ which takes on its values in
$]0 \, , {\pi}[$ and that the scattering by ${\rm N}$ pointlike
scatterers is described by a system of only ${\rm N}$ equations. We then show with the help 
of numerical calculations that there are two stages in the variation of the mean cross 
section as the density of the samples (the radius of the target) increases (decreases). 
Depending on the value of ${\delta}$,
the mean cross section first either increases or decreases, each one of the two behaviours 
being originated by double scattering; it decreases uniformly for any value of ${\delta}$ as the 
density increases further on, a behaviour which results from multiple scattering
and which follows that of the cross section for 
diffusion by a hard sphere potential of decreasing radius. The 
expression of the mean cross section is derived in the particular case of an unlimited 
number of contributions of successive scatterings. 
\end{abstract}

\maketitle

\section{Introduction}

Many works (see, e.g., \cite{rn},~\cite{vcl},~\cite{lt},~\cite{ts},~\cite{am}) have been done 
concerning the description of the quantum scattering of a 
particle in disordered systems (one usually prefers to speak of "transport" rather 
than "scattering" if the system is infinite or semi-infinite), but all of them fail to 
preserve unitarity. The lack of unitarity has two different origins:

$\rightarrow$ the most evident origin lies in the fact that not all sequences of multiple 
scattering are taken into account, but usually only those in which the particle is only scattered once by 
each particular scatterer 

$\rightarrow$ another origin for the lack of unitarity lies in the fact that the operator which describes  
the scattering by a single scatterer is always replaced by its first-order Born approximation. 

The paper has two purposes, which are:

\point to develop a formalism in which unitarity is preserved, which means 1) to use the full operator describing
the scattering by a single scatterer and 2) to take into account all sequences of multiple scattering 

\point \point to study scattering "by" (instead of "in") disordered systems 1) of finite size (which is still 
largely a new topic, see however~\cite{ln}), and 2) consisting of individual scatterers, the scattering by which 
is characterized by phase shifts (disordered systems are usually modelled by a continuous random 
potential "landscape"), in order
to find quantum effects which cannot be seen in disordered systems of infinite size.

These two goals can be simultaneously achieved by modelling disordered systems as a target with a random distribution
of pointlike scatterers.

The paper is divided into six sections and has two appendices. Section~2 gives the 
derivation of the infinite set of equations which describes the quantum scattering of a 
particle by a target consisting of identical potential wells with spherical symmetry. In
Section 3 we prove that scattering by a pointlike scatterer is characterized by a single
phase shift which takes on its values in the interval $]0 \, , {\pi}[$ and
that the infinite set of equations reduces to a system of only ${\rm N}$ equations if the target 
consists of ${\rm N}$ pointlike scatterers. In Section 4 we show that the optical theorem holds
for any number of pointlike scatterers. In Section 5 we present and discuss the results
of a numerical study about the scattering by a disordered target consisting of pointlike
scatterers, in which we have examined the variation of the mean cross section with the
density of the samples, and so with the radius of the target, for different values of the
phase shift. A summary of the main results of the paper is given in Section 6. Each of
the appendices is devoted to the derivation of an expression of the mean cross section
which has been used in the interpretation of the numerical results, namely that in the 
approximation of single and double scattering (Appendix A) and that for an unlimited 
number of contributions of successive scatterings (Appendix B).  

\section{Formalism}

Let us consider the scattering of a quantum particle by a target which consists of a set
of ${\rm N}$ stationary scatterers. Each scatterer is modelled by a potential well, the potential 
function being chosen square integrable if the depth is not finite. The scatterers are so 
arranged that no two neighbouring wells are close enough to overlap. The Hamiltonian 
operator of the system is given by   
\beq
  {\hat H} = {\hat h}_{0} + \sum_{r=1}^{\rm N} {\hat v}_{r},
\label{e1}
\eeq
where ${\hat h}_{0}$ is the kinetic energy operator and ${\hat v}_{r}$ is the operator of
multiplication by the potential function of the $r$th scatterer. The particle, which has mass $m$, is assumed to move     
with momentum ${\hbar}{\vec k}$ before scattering; it therefore has the energy
$E = {{\hbar}^{2}}{{k}^{2}}/2m$, where $k$ is the magnitude of the wave vector.

The operators which describe the scattering by a single scatterer and by the whole set of scatterers 
are respectively given by~\cite{gw}
\beq
  {\hat t}_{r}(E) = {\hat v}_{r} + {\hat v}_{r} 
  \hskip -0.028 truecm
  \frac{1}{(E + i{\epsilon}){\hat I} - {\hat h}_{0} - {\hat v}_{r}} \hskip +0.010 truecm {\hat v}_{r} 
  \, \, \, \, \, \, \, \, \, \, \, \, \, \, \, \, \, \, \, \, \, \, \, \, \, \, \, \, \, \, \, \, \, \, \, \, 
  \hskip -0.019 truecm (r = 1,...,{\rm N}),
\label{e2}
\eeq
and
\beq
  {\hat T}(E) = \Biggl ( \hskip +0.050 truecm \sum_{r=1}^{\rm N} {\hat v}_{r} \hskip -0.065 truecm \Biggr )
  \hskip -0.025 truecm
  + \hskip -0.025 truecm \Biggl ( \hskip +0.050 truecm \sum_{r=1}^{\rm N} {\hat v}_{r} \hskip -0.065 truecm \Biggr )
  \hskip -0.038 truecm
  \frac{1}{(E + i{\epsilon}){\hat I} - {\hat H}} 
  \hskip -0.038 truecm
  \Biggl ( \hskip +0.050 truecm \sum_{s=1}^{\rm N} {\hat v}_{s} \hskip -0.065 truecm \Biggr ) \hskip -0.030 truecm .
\label{e3}
\eeq
These operators are integral operators in the momentum representation; each kernel is bounded 
and (uniformly) continuous, as the corresponding potential function is square integrable and has  
bounded support.

It is convenient to write the operator ${\hat T}(E)$ as
\beq
  {\hat T}(E) = \sum_{r=1}^{\rm N} {\hat T}_{r}(E), 
\label{e4}
\eeq
with
\beq
  {\hat T}_{r}(E) = {\hat v}_{r} 
  + \sum_{s=1}^{\rm N} {\hat v}_{r} \hskip -0.028 truecm \frac{1}{(E + i{\epsilon}){\hat I} - {\hat H}} 
  \hskip +0.010 truecm {\hat v}_{s}.
\label{e5}
\eeq
An operator ${\hat T}_{r}(E)$ may be considered as the transition operator corresponding to the 
(hypothetical) scattering process in which the particle is always last scattered by the $r$th scatterer. These  
operators are related by the set of equations~\cite{ajp}
\begin{eqnarray}
  {\hat T}_{r}(E) && = {\hat t}_{r}(E) + 
  \hskip -0.11 truecm
  \sum_{\scriptstyle s=1 \atop \scriptstyle (s \neq r)}^{\rm N} 
  \hskip -0.11 truecm 
  {\hat t}_{r}(E) \frac{1}{(E + i{\epsilon}){\hat I} - {\hat h}_{0}} {\hat T}_{s}(E)   
  \nonumber\\
  && \, \, \, \, \, \, \, \, \, \, \, \, \, \, \, \, \, \, \, \, \, \, \, \, \, \, \, \, \, \, \, \, \, \, \,
  \, \, \, \, \, \, \, \, \, \, \, \, \, \, \, \, \, \, \, \, \, \, \, \, \, \, \, \, \, \, \, \, \, \, \, \,
  \, \, \, \, \, \, \, \, \, \, \, \, \, \, \, \, \, \, \, \, \, \, \, \, \, \, \, \, \, \, \, \, \, \, \, \, 
  \, \, \, \, \, \, \, \, \hskip +0.023 truecm (r = 1,...,{\rm N}).
\label{e6}
\end{eqnarray}
Repeated substitution of each equation into the others shows that the particle can be scattered any number 
of times by each particular scatterer; scattering by the target may therefore be considered the result of  
multiple scattering among the scatterers~\cite{w}. 
 
The probability amplitude $f({\vec k}^{'} \hskip -0.040 truecm ;{\vec k})$ for the particle to be scattered in the
direction of the wave vector ${\vec k}^{'}$ is proportional to the particular value 
$T({\vec k}^{'} \hskip -0.040 truecm ,{\vec k} \, ;E)$ of the kernel of the operator ${\hat T}(E)$~\cite{s}. We have 
\beq
  f({\vec k}^{'} \hskip -0.040 truecm ;{\vec k}) 
  = - {(2 \pi)}^{2} \, \hskip -0.019 truecm \frac{m}{{\hbar}^{2}} \, T({\vec k}^{'} \hskip -0.040 truecm ,{\vec k} \, ;E)
  = - {(2 \pi)}^{2} \, \hskip -0.019 truecm \frac{m}{{\hbar}^{2}} \sum_{r=1}^{\rm N} T_{r}({\vec k}^{'} \hskip -0.040 truecm ,{\vec k} \, ;E),
\label{e7}
\eeq
where $T_{r}({\vec k}^{'} \hskip -0.040 truecm ,{\vec k} \, ;E)$ is the value of the kernel of the operator 
${\hat T}_{r}(E)$ for the chosen wave vectors. It follows from~(\ref{e6}) that the kernels of these operators 
are related by the set of equations  
\begin{eqnarray}
  T_{r}({\vec \eta} \, ,{\vec \xi} \, ;E) && = t_{r}({\vec \eta} \, ,{\vec \xi} \, ;E)
  + \frac{2m}{{\hbar}^{2}} 
  \hskip -0.11 truecm
  \sum_{\scriptstyle s=1 \atop \scriptstyle (s \neq r)}^{\rm N} 
  \hskip -0.11 truecm  
  \int {\d {\vec \kappa}} \, \frad{t_{r}({\vec \eta} \, ,{\vec \kappa} \, ;E) \,              
  T_{s}({\vec \kappa} \, ,{\vec \xi} \, ;E)}{k^{2} - {\kappa}^{2} + i{\epsilon}} 
  \nonumber\\
  && \, \, \, \, \, \, \, \, \, \, \, \, \, \, \, \, \, \, \, \, \, \, \, \, \, \, \, \, \, \, \, \, \, \, \, \,
  \, \, \, \, \, \, \, \, \, \, \, \, \, \, \, \, \, \, \, \, \, \, \, \, \, \, \, \, \, \, \, \, \, \, \, \, \,
  \, \, \, \, \, \, \, \, \, \, \, \, \, \, \, \, \, \, \, \, \, \, \, \, \, \, \, \, \,
  \hskip +0.008 truecm (r = 1,...,{\rm N}),               
\label{e8}
\end{eqnarray}
where $t_{r}({\vec \eta} \, ,{\vec \xi} \, ;E)$ is the kernel of the operator ${\hat t}_{r}(E)$ and ${\kappa}$ 
is the magnitude of the wave vector ${\vec \kappa}$. In particular, 
\begin{eqnarray}
  T_{r}({\vec k}^{'} \hskip -0.040 truecm ,{\vec k} \, ;E) && = t_{r}({\vec k}^{'} \hskip -0.040 truecm ,{\vec k} \, ;E) 
  + \frac{2m}{{\hbar}^{2}} 
  \hskip -0.11 truecm
  \sum_{\scriptstyle s=1 \atop \scriptstyle (s \neq r)}^{\rm N} 
  \hskip -0.11 truecm
  \int {\d {\vec \kappa}} \, \frad{t_{r}({\vec k}^{'} \hskip -0.040 truecm ,{\vec \kappa} \, ;E) \,              
  T_{s}({\vec \kappa} \, ,{\vec k} \, ;E)}{k^{2} - {\kappa}^{2} + i{\epsilon}} 
  \nonumber\\
  && \, \, \, \, \, \, \, \, \, \, \, \, \, \, \, \, \, \, \, \, \, \, \, \, \, \, \, \, \, \, \, \, \, \, \, \,
  \, \, \, \, \, \, \, \, \, \, \, \, \, \, \, \, \, \, \, \, \, \, \, \, \, \, \, \, \, \, \, \, \, \, \, \, \,
  \, \, \, \, \, \, \, \, \, \, \, \, \, \, \, \, \, \, \, \, \, \, \, \, \, \, \, \, \,  
  \hskip -0.031 truecm (r = 1,...,{\rm N}).    
\label{e9}
\end{eqnarray}

We shall assume that the scatterers are identical and restrict ourselves to potential functions with 
spherical symmetry. The kernel of an operator ${\hat t}_{r}(E)$ can then be written in the form
\beq
  t_{r}({\vec \eta} \, ,{\vec \xi} \, ;E) = \e^{ \, i ({\vec \xi} - {\vec \eta}).{\vec R}_{r}} \, 
  t({\vec \eta} \, ,{\vec \xi} \, ;E) \, \, \, \, \, \, \, \, \, \, \, \, \, \, \, \, \, \, \, \, \, \, \,  
  \, \, \, \, \, \, \, \, \, \, \, \, \, \, \, \, \, \, \, \, \, \, 
  \hskip -0.030 truecm (r = 1,...,{\rm N}),
\label{e10}
\eeq
where $t({\vec \eta} \, ,{\vec \xi} \, ;E)$ is the function to which any such kernel would reduce if the centre of 
the corresponding potential well were taken as the origin of coordinates and ${\vec R}_{r}$ is the vector from the
chosen origin to the centre of the $r$th potential well.

Substituting this equation into~(\ref{e9}) and writing each function 
$T_{r}({\vec \kappa} \, ,{\vec k} \, ;E)$ in the form   
\beq
  T_{r}({\vec \kappa} \, ,{\vec k} \, ;E) = \e^{ \, -i {\vec \kappa}.{\vec R}_{r}} 
  \sum_{u=1}^{\rm N} \e^{ \, i{\vec k}.{\vec R}_{u}} \, {\cal T}_{ru}({\vec \kappa} \, ,{\vec k} \, ;E) 
  \, \, \, \, \, \, \, \, \, \, \, \, \, \, \, \, \,  
  \hskip +0.005 truecm (r = 1,...,{\rm N})       
\label{e11}
\eeq  
in~(\ref{e7}) and~(\ref{e9}), we obtain
\beq
  f({\vec k}^{'} \hskip -0.040 truecm ;{\vec k}) 
  = - {(2 \pi)}^{2} \, \hskip -0.019 truecm \frac{m}{{\hbar}^{2}} \sum_{r = 1}^{\rm N} \sum_{u=1}^{\rm N}
  \e^{ \, -i {{\vec k}^{'}} \hskip -0.095 truecm .{\vec R}_{r}} \, \e^{ \, i {\vec k}.{\vec R}_{u}} \, 
  {\cal T}_{ru}({\vec k}^{'} \hskip -0.040 truecm ,{\vec k} \, ;E),
\label{e12}
\eeq
with 
\begin{eqnarray}
  {\cal T}_{ru}({\vec k}^{'} \hskip -0.040 truecm ,{\vec k} \, ;E) 
  && = t({\vec k}^{'} \hskip -0.040 truecm ,{\vec k} \, ;E) \, {\delta}_{ru}  
  \nonumber\\
  && \, \, \, + \frac{2m}{{\hbar}^{2}} 
  \hskip -0.11 truecm
  \sum_{\scriptstyle s=1 \atop \scriptstyle (s \neq r)}^{\rm N} 
  \hskip -0.11 truecm
  \int {\d {\vec \kappa}} \, \frad{\e^{ \, i {\vec \kappa}.({\vec R}_{r} - {\vec R}_{s})} \,
  t({\vec k}^{'} \hskip -0.040 truecm ,{\vec \kappa} \, ;E) \,
  {\cal T}_{su}({\vec \kappa} \, ,{\vec k} \, ;E)}{k^{2} - {\kappa}^{2} + i{\epsilon}}.                    
  \nonumber\\
\label{e13}
\end{eqnarray}                                         

The new set of equations can be transformed to another with no dependence upon direction. This is done by 
expanding both sides of each equation in the basis 
$\lbrace Y_{l m} \rbrace$ of the spherical harmonics and comparing the two expansions term by term. 

The expansion of the functions ${\cal T}_{ru}({\vec \kappa} \, ,{\vec k} \, ;E)$ is given by
\begin{eqnarray}
  && {\cal T}_{ru}({\vec \kappa} \, ,{\vec k} \, ;E)
  \nonumber\\
  && = {4 \pi} \sum_{l_{1} \hskip -0.015 truecm =0}^{\infty} \sum_{m_{1} \hskip -0.015 truecm =-l_{1}}^{l_{1}} 
  \sum_{l_{2}=0}^{\infty} \sum_{m_{2}=-l_{2}}^{l_{2}}   
  {\cal T}_{l_{1} \hskip -0.025 truecm m_{1} \hskip -0.021 truecm ; \hskip +0.023 truecm l_{2} m_{2}}^{\, (r u)}({\kappa} \, ,k \, ;E) \,
  Y_{l_{1} \hskip -0.025 truecm m_{1}}^{*} \hskip -0.015 truecm ({\Omega}_{{\vec \kappa}}) \, 
  Y_{l_{2} m_{2}}({\Omega}_{{\vec k}})
  \nonumber\\
  && \, \, \, \, \, \, \, \, \, \, \, \, \, \, \, \, \, \, \, \, \, \, \, \, \, \, \, \, \, \, \, \, \, \, \, \, \, 
  \, \, \, \, \, \, \, \, \, \, \, \, \, \, \, \, \, \, \, \, \, \, \, \, \, \, \, \, \, \, \, \, \, \, \, \, \, \, 
  \, \, \, \, \, \, \, \, \, \, \, \, \, \, \, \, \, \, \, \, \, \, \, \, \, \, \, \, \, \, \, \, \, \, \, \, \, \, 
  \, \, \, \, \, \, \, \, \, 
  \hskip -0.019 truecm (r,u = 1,...,{\rm N}),
\label{e14}
\end{eqnarray} 
where ${\Omega}_{{\vec \kappa}}$ and ${\Omega}_{{\vec k}}$ are the directions of the 
wave vectors. Since 
the potential function is invariant under rotation, the function 
$t({\vec k}^{'} \hskip -0.040 truecm ,{\vec \kappa} \, ;E)$ has the simpler expansion
\beq
  t({\vec k}^{'} \hskip -0.040 truecm ,{\vec \kappa} \, ;E) 
  = {4 \pi} \sum_{l=0}^{\infty} \sum_{m=-l}^{l} t_{l}(k \, ,{\kappa} \, ;E) \,          
  Y_{l m}^{*}({\Omega}_{{\vec k}^{'}} \hskip -0.058 truecm ) \, Y_{l m}({\Omega}_{{\vec \kappa}}),
\label{e15}
\eeq 
where ${\Omega}_{{\vec k}^{'}}$ is the considered direction of scattering. The functions
$\e^{ \, i {\vec \kappa}.({\vec R}_{r} - {\vec R}_{s})}$ must also be expanded to obtain the expanded form of the 
right-hand side of the equations. One has~\cite{j}                              
\begin{eqnarray}
  && \e^{ \, i {\vec \kappa}.({\vec R}_{r} - {\vec R}_{s})} = {4 \pi} \sum_{l=0}^{\infty} \sum_{m=-l}^{l} i^{l} \,
  j_{l}({\kappa} R_{rs}) \, Y_{l m}^{*}({\Omega}_{{\vec \kappa}}) \, Y_{l m}({\Omega}_{{\vec R}_{sr}})
  \nonumber\\
  && \, \, \, \, \, \, \, \, \, \, \, \, \, \, \, \, \, \, \, \, \, \, \, \, \, \, \, \, \, \, \, \, \, \, \, \, \, 
  \, \, \, \, \, \, \, \, \, \, \, \, \, \, \, \, \, \, \, \, \, \, \, \, \, \, \, \, \, \, \, \, \, \, \, \, \, \, 
  \, \, \, \, \, \, \, \, \, \, \, \, \, \, \, \, \, \, \, \, \, \, \, \, \, \, \, \,    
  \hskip -0.056 truecm (r,s = 1,...,{\rm N} \, \, ; \, \, s \neq r),
\label{e16}
\end{eqnarray}
where $R_{rs}$ and ${\Omega}_{{\vec R}_{sr}}$ are the magnitude and direction of the 
vector ${\vec R}_{sr} = {\vec R}_{r} - {\vec R}_{s}$ from the centre of the 
$s$th to that of the $r$th potential well and $j_{l}(x)$ is the spherical Bessel function of order $l$. 

Substitution of these expansions into~(\ref{e13}) and use of the orthonormality relations of  
the spherical harmonics leads to the infinite set of equations  
\begin{eqnarray}
  && {\cal T}_{l_{1} \hskip -0.025 truecm m_{1} \hskip -0.021 truecm ; \hskip +0.023 truecm l_{2} m_{2}}^{\, (r u)}(k \, ,k \, ;E)
  = t_{l_{1}} \hskip -0.015 truecm (k \, ,k \, ;E) \, {\delta}_{r u} \, 
  {\delta}_{l_{1} \hskip -0.025 truecm l_{2}} \hskip +0.040 truecm 
  {\delta}_{m_{1} \hskip -0.025 truecm m_{2}}
  \nonumber\\
  && \, \, \, + {8 \pi} \, \frac{m}{{\hbar}^{2}} 
  \sum_{l_{3}=0}^{\infty} \sum_{m_{3}=-l_{3}}^{l_{3}} \sum_{l_{4}=0}^{\infty} \sum_{m_{4}=-l_{4}}^{l_{4}}
  i^{l_{4}} \, {\sqrt{\frac{{4 \pi} (2l_{3} + 1) (2l_{4} + 1)}{2l_{1} + 1}}}  
  \nonumber\\  
  && \, \, \, \, \, \, \, \, \times \left \langle \, l_{3} \, l_{4} \, 0 \, 0 \, | \, l_{1} \, \hskip -0.025 truecm 0 \, \right \rangle   
  \left \langle \, l_{3} \, l_{4} \, m_{3} \, m_{4} \, | \, l_{1} \, \hskip -0.025 truecm m_{1} \, \hskip -0.035 truecm
  \right \rangle 
  \nonumber\\ 
  && \, \, \, \, \, \, \, \, \times 
  \hskip -0.11 truecm
  \sum_{\scriptstyle s=1 \atop \scriptstyle (s \neq r)}^{\rm N} 
  \hskip -0.11 truecm
  Y_{l_{4} m_{4}}({\Omega}_{{\vec R}_{sr}})      
  \int_{0}^{\infty} {{\kappa}^{2} \d {\kappa}} \, \frad{j_{l_{4}}({\kappa} R_{rs}) \,
  t_{l_{1}} \hskip -0.015 truecm (k \, ,{\kappa} \, ;E) \, 
  {\cal T}_{l_{3} m_{3} ; \hskip +0.023 truecm l_{2} m_{2}}^{\, (s u)}({\kappa} \, ,k \, ;E)}
  {k^{2} - {\kappa}^{2} + i{\epsilon}} 
  \nonumber\\ 
  && \, \, \, \, \, \, \, \, \, \, \, \, \, \, \, \, \, \, \, \, \, \, \, \, \, \, \, \, \, \, \, \, \, \, \, \, \, \, \, \,   
  \, \, \, \, \, \, \, \, \, \, \, \, \, \, \, \, \, \, \, \, (r,u = 1,...,{\rm N})
  \nonumber\\
  && \, \, \, \, \, \, \, \, \, \, \, \, \, \, \, \, \, \, \, \, \, \, \, \, \, \, \, \, \, \, \, \, \, \, \, \, \, \, \, \,
  \, \, \, \, \, \, \, \, \, \, \, \, \, \, \, \, \, \, \, \, (l_{1} \hskip -0.015 truecm ,l_{2} = 0,1,... \, \, ; \, \, 
  m_{1 \hskip -0.015 truecm (2)} = -l_{1 \hskip -0.015 truecm (2)},...,l_{1 \hskip -0.015 truecm (2)}), 
\label{e17}
\end{eqnarray}
where $\left \langle \, l_{3} \, l_{4} \, 0 \, 0 \, | \, l_{1} \, \hskip -0.025 truecm 0 \, \right \rangle$ and 
$\left \langle \, l_{3} \, l_{4} \, m_{3} \, m_{4} \, | \, l_{1} \, \hskip -0.025 truecm m_{1} \, \hskip -0.035 truecm
\right \rangle$ are Clebsch-Gordan 
coefficients~\cite{s} coming from the integration over ${\Omega}_{{\vec \kappa}}$. Since the coefficient 
$\left \langle \, l_{3} \, l_{4} \, 0 \, 0 \, | \, l_{1} \, \hskip -0.025 truecm 0 \, \right \rangle$ is equal 
to $0$ when the sum $l_{1} + l_{3} + l_{4}$ is odd, the right-hand side of the equations  
includes only the terms for which this sum is even.   

The integral in~(\ref{e17}) may be evaluated by contour integration. The result is   
\begin{eqnarray}      
  && 2 \int_{0}^{\infty} {{\kappa}^{2} \d {\kappa}} \, \frad{j_{l_{4}}({\kappa} R_{rs}) \,
  t_{l_{1}} \hskip -0.015 truecm (k \, ,{\kappa} \, ;E) \, 
  {\cal T}_{l_{3} m_{3} ; \hskip +0.023 truecm l_{2} m_{2}}^{\, (s u)}({\kappa} \, ,k \, ;E)}
  {k^{2} - {\kappa}^{2} + i{\epsilon}} 
  \nonumber\\ 
  && = \int_{-\infty}^{\infty} {{\kappa}^{2} \d {\kappa}} \, \frad{h^{(1)}_{l_{4}}({\kappa} R_{rs}) \,
  t_{l_{1}} \hskip -0.015 truecm (k \, ,{\kappa} \, ;E) \, 
  {\cal T}_{l_{3} m_{3} ; \hskip +0.023 truecm l_{2} m_{2}}^{\, (s u)}({\kappa} \, ,k \, ;E)}
  {k^{2} - {\kappa}^{2} + i{\epsilon}}   
  \nonumber\\
  && = -i {\pi} k \, h^{(1)}_{l_{4}}(k R_{rs}) \, t_{l_{1}} \hskip -0.015 truecm (k \, ,k \, ;E) \,
  {\cal T}_{l_{3} m_{3} ; \hskip +0.023 truecm l_{2} m_{2}}^{\, (s u)}(k \, ,k \, ;E)     
  \nonumber\\
  && \, \, \, \, \, \, \, \, \, \, \, \, \, \, \, \, \, \, \, \, \, \, \, \, \, \, \, \, \, \, \, \, \, \, \, \, \, \,     
  \, \, \, \, \, \, \, \, (r,s,u = 1,...,{\rm N} \, \, ; \, \, s \neq r) 
  \nonumber\\
  && \, \, \, \, \, \, \, \, \, \, \, \, \, \, \, \, \, \, \, \, \, \, \, \, \, \, \, \, \, \, \, \, \, \, \, \, \, \,   
  \, \, \, \, \, \, \, \, (l_{1} \hskip -0.015 truecm ,l_{2},l_{3},l_{4} = 0,1,... \, \, ; \, \, m_{2(3)} = -l_{2(3)},...,l_{2(3)}),
\label{e18}
\end{eqnarray}
where $h^{(1)}_{l}(x)$ is the spherical Hankel function of the first kind of order $l$. The 
range of integration could be extended over the whole line by using the relations 
\beq
  t_{l_{1}} \hskip -0.015 truecm (k \, ,-{\kappa} \, ;E) = (- 1)^{l_{1}} \, t_{l_{1}} \hskip -0.015 truecm (k \, ,{\kappa} \, ;E)
\label{e19}
\eeq 
and 
\beq
  {\cal T}_{l_{3} m_{3} ; \hskip +0.023 truecm l_{2} m_{2}}^{\, (s u)}(-{\kappa} \, ,k \, ;E)
  = (- 1)^{l_{3}} \, {\cal T}_{l_{3} m_{3} ; \hskip +0.023 truecm l_{2} m_{2}}^{\, (s u)}({\kappa} \, ,k \, ;E), 
\label{e20}
\eeq 
which follow from the rotational invariance of the potential function, and  
the fact that the sum $l_{1} + l_{3} + l_{4}$ is even. The
contribution of the complex part of the contour to the integral is
arbitrarily small, as can be shown by using the inequalities
\beq
  \abs{t_{l_{1}} \hskip -0.015 truecm (k \, ,\Re \lbrace {\kappa} \rbrace + i \Im \lbrace {\kappa} \rbrace \, ;E)} 
  < A_{l_{1}} \, \hskip -0.015 truecm \e^{ \, + \abs{\Im \lbrace {\kappa} \rbrace} a}
\label{e21}
\eeq
and
\beq
  \abs{{\cal T}_{l_{3} m_{3} ; \hskip +0.023 truecm l_{2} m_{2}}^{\, (s u)}(\Re \lbrace {\kappa} \rbrace 
  + i \Im \lbrace {\kappa} \rbrace \, ,k \, ;E)} < B_{l_{2} l_{3}} \, \e^{ \, + \abs{\Im \lbrace {\kappa} \rbrace} a},  
\label{e22}
\eeq
where $a$ is the radius of the potential wells ($A_{l_{1}}$ and $B_{l_{2} l_{3}}$ are constants), and         
the fact that any distance $R_{rs}$ is greater than $2a$.

Substituting~(\ref{e18}) into~(\ref{e17}), we obtain the set of equations which describes the 
scattering of a particle by a target consisting of identical potential wells with spherical 
symmetry. It is 
\begin{eqnarray}
  && {\cal T}_{l_{1} \hskip -0.025 truecm m_{1} \hskip -0.021 truecm ; \hskip +0.023 truecm l_{2} m_{2}}^{\, (r u)}(k \, ,k \, ;E)
  \nonumber\\
  && \, \, \, + i \, {(2 \pi)^{2}} \, \hskip -0.019 truecm \frac{m k}{{\hbar}^{2}} \, 
  t_{l_{1}} \hskip -0.015 truecm (k \, ,k \, ;E)
  \sum_{l_{3}=0}^{\infty} \sum_{m_{3}=-l_{3}}^{l_{3}} \sum_{l_{4}=0}^{\infty} \sum_{m_{4}=-l_{4}}^{l_{4}}
  i^{l_{4}} \, {\sqrt{\frac{{4 \pi} (2l_{3} + 1) (2l_{4} + 1)}{2l_{1} + 1}}}  
  \nonumber\\
  && \, \, \, \, \, \, \, \, \times \left \langle \, l_{3} \, l_{4} \, 0 \, 0 \, | \, l_{1} \, \hskip -0.025 truecm 0 \, \right \rangle  
  \left \langle \, l_{3} \, l_{4} \, m_{3} \, m_{4} \, | \, l_{1} \, \hskip -0.025 truecm m_{1} \, \hskip -0.035 truecm
  \right \rangle 
  \nonumber\\ 
  && \, \, \, \, \, \, \, \, \times 
  \hskip -0.11 truecm
  \sum_{\scriptstyle s=1 \atop \scriptstyle (s \neq r)}^{\rm N}
  \hskip -0.11 truecm
  h^{(1)}_{l_{4}}(k R_{rs}) \,
  Y_{l_{4} m_{4}}({\Omega}_{{\vec R}_{rs}}) \, 
  {\cal T}_{l_{3} m_{3} ; \hskip +0.023 truecm l_{2} m_{2}}^{\, (s u)}(k \, ,k \, ;E)   
  \nonumber\\
  && = t_{l_{1}} \hskip -0.015 truecm (k \, ,k \, ;E) \, {\delta}_{r u} \, 
  {\delta}_{l_{1} \hskip -0.025 truecm l_{2}} \hskip +0.040 truecm 
  {\delta}_{m_{1} \hskip -0.025 truecm m_{2}}
  \nonumber\\
  && \, \, \, \, \, \, \, \, \, \, \, \, \, \, \, \, \, \, \, \, \, \, \, \, \, \, \, \, \, \, \, \, \, \, \, \, \, \, \, \,   
  \, \, \, \, \, \, \, \, \, \, \, \, \, \, \, \, \, \, \, \, \, 
  (r,u = 1,...,{\rm N})
  \nonumber\\
  && \, \, \, \, \, \, \, \, \, \, \, \, \, \, \, \, \, \, \, \, \, \, \, \, \, \, \, \, \, \, \, \, \, \, \, \, \, \, \, \, 
  \, \, \, \, \, \, \, \, \, \, \, \, \, \, \, \, \, \, \, \, \, 
  (l_{1} \hskip -0.015 truecm ,l_{2} = 0,1,... \, \, ; \, \, 
  m_{1 \hskip -0.015 truecm (2)} = -l_{1 \hskip -0.015 truecm (2)},...,l_{1 \hskip -0.015 truecm (2)}).
\label{e23}
\end{eqnarray}  

\section{Pointlike scatterers}               

The set has infinitely many equations because the expansion of the function
$t({\vec k}^{'} \hskip -0.040 truecm ,{\vec k} \, ;E)$ 
has an unlimited number of terms. However, in the particular case in which
$t({\vec k}^{'} \hskip -0.040 truecm ,{\vec k} \, ;E)$ does not depend on the directions of the    
wave vectors, which means that  
\beq           
  t({\vec k}^{'} \hskip -0.040 truecm ,{\vec k} \, ;E) = t_{0}(k \, ,k \, ;E),
\label{e24}
\eeq  
it turns out as a consequence of the invariance of the system under time reversal that only the ${\rm N}^{2}$ 
expansion coefficients 
${\cal T}_{0 0 \hskip +0.023 truecm ; \hskip +0.023 truecm 0 0}^{\, (r u)}(k \, ,k \, ;E)$
are different from $0$ and so the set has only ${\rm N}^{2}$
equations. The proof is as follows. Substitution of~(\ref{e24}), which may be written 
$t_{l}(k \, ,k \, ;E) = t_{0}(k \, ,k \, ;E) \, {\delta}_{l \hskip +0.015 truecm 0}$, into~(\ref{e23}) leads to 
the equality 
\begin{eqnarray} 
  && {\cal T}_{l_{1} \hskip -0.025 truecm m_{1} \hskip -0.021 truecm ; \hskip +0.023 truecm l_{2} m_{2}}^{\, (r u)}(k \, ,k \, ;E) =
  {\cal T}_{0 0 \hskip +0.023 truecm ; \hskip +0.023 truecm l_{2} m_{2}}^{\, (r u)}(k \, ,k \, ;E) \, 
  {\delta}_{l_{1} \hskip -0.025 truecm 0} 
  \nonumber\\
  && \, \, \, \, \, \, \, \, \, \, \, \, \, \, \, \, \, \, \, \, \, \, \, \, \, \, \, \, \, \, \, \, \, \, \, \, \, \,    
  \, \, \, \, \, \, \, \, \, \, \, \, \, \, \, \, \, \, \, \, \, \, \, \, \,
  (r,u = 1,...,{\rm N})  
  \nonumber\\
  && \, \, \, \, \, \, \, \, \, \, \, \, \, \, \, \, \, \, \, \, \, \, \, \, \, \, \, \, \, \, \, \, \, \, \, \, \, \,  
  \, \, \, \, \, \, \, \, \, \, \, \, \, \, \, \, \, \, \, \, \, \, \, \, \,
  (l_{1} \hskip -0.015 truecm ,l_{2} = 0,1,... \, \, ; \, \, 
  m_{1 \hskip -0.015 truecm (2)} = -l_{1 \hskip -0.015 truecm (2)},...,l_{1 \hskip -0.015 truecm (2)}).
\label{e25}
\end{eqnarray}                                                         
Time-reversal symmetry implies that  
\beq
  f({\vec k}^{'} \hskip -0.040 truecm ;{\vec k}) = f(-{\vec k} \, ;-{\vec k}^{'} \hskip -0.053 truecm ).
\label{e26}
\eeq                              
It follows that the functions ${\cal T}_{ru}({\vec k}^{'} \hskip -0.040 truecm ,{\vec k} \, ;E)$ satisfy the relation
\beq
  {\cal T}_{ru}({\vec k}^{'} \hskip -0.040 truecm ,{\vec k} \, ;E) 
  = {\cal T}_{ur}(-{\vec k} \, ,-{\vec k}^{'} \hskip -0.040 truecm ;E)
  \, \, \, \, \, \, \, \, \, \, \, \, \, \, \, \, \, \, \, \, \, \, \, \, \, \, \, \, \, \, \, \, 
  \, \, \, \, \, \, \, \, \, \, \, \, \, \hskip +0.018 truecm (r,u = 1,...,{\rm N}),
\label{e27}
\eeq 
and so their expansion coefficients the relation
\begin{eqnarray}
  && {\cal T}_{l_{2} m_{2} ; \hskip +0.023 truecm l_{1} \hskip -0.025 truecm m_{1}}^{\, (r u)} \hskip -0.015 truecm (k \, ,k \, ;E) =
  (-1)^{l_{1} \hskip -0.025 truecm - \hskip +0.014 truecm m_{1} + \hskip +0.025 truecm l_{2} - \hskip +0.014 truecm m_{2}} \, 
  {\cal T}_{l_{1} \hskip -0.025 truecm -m_{1} \hskip -0.021 truecm ; \hskip +0.023 truecm l_{2} -m_{2}}^{\, (u r)}(k \, ,k \, ;E)  
  \nonumber\\
  && \, \, \, \, \, \, \, \, \, \, \, \, \, \, \, \, \, \, \, \, \, \, \, \, \, \, \, \, \, \, \, \, \, \, \, \, \, \,    
  \, \, \, \, \, \, \, \, \, \, \, \, \, \, \, \, \, \, \, \, \, \, \, \, \,
  (r,u = 1,...,{\rm N})  
  \nonumber\\
  && \, \, \, \, \, \, \, \, \, \, \, \, \, \, \, \, \, \, \, \, \, \, \, \, \, \, \, \, \, \, \, \, \, \, \, \, \, \,  
  \, \, \, \, \, \, \, \, \, \, \, \, \, \, \, \, \, \, \, \, \, \, \, \, \,
  (l_{1} \hskip -0.015 truecm ,l_{2} = 0,1,... \, \, ; \, \, 
  m_{1 \hskip -0.015 truecm (2)} = -l_{1 \hskip -0.015 truecm (2)},...,l_{1 \hskip -0.015 truecm (2)}).
\label{e28}
\end{eqnarray}    
Alternate use of~(\ref{e25}) and~(\ref{e28}) then gives  
\begin{eqnarray}
  {\cal T}_{l_{1} \hskip -0.025 truecm m_{1} \hskip -0.021 truecm ; \hskip +0.023 truecm l_{2} m_{2}}^{\, (r u)}(k \, ,k \, ;E) && =
  {\cal T}_{0 0 \hskip +0.023 truecm ; \hskip +0.023 truecm l_{2} m_{2}}^{\, (r u)}(k \, ,k \, ;E) \, 
  {\delta}_{l_{1} \hskip -0.025 truecm 0} 
  \nonumber\\
  && = (-1)^{l_{2} - \hskip +0.014 truecm m_{2}} \, {\cal T}_{l_{2} -m_{2} ; \hskip +0.023 truecm 0 0}^{\, (u r)}(k \, ,k \, ;E) \, 
  {\delta}_{l_{1} \hskip -0.025 truecm 0}
  \nonumber\\ 
  && = {\cal T}_{0 0 \hskip +0.023 truecm ; \hskip +0.023 truecm 0 0}^{\, (u r)}(k \, ,k \, ;E) \,
  {\delta}_{l_{1} \hskip -0.025 truecm 0} \, {\delta}_{l_{2} 0}  
  \nonumber\\   
  && = {\cal T}_{0 0 \hskip +0.023 truecm ; \hskip +0.023 truecm 0 0}^{\, (r u)}(k \, ,k \, ;E) \,
  {\delta}_{l_{1} \hskip -0.025 truecm 0} \, {\delta}_{l_{2} 0} 
  \nonumber\\    
  && \, \, \, \, \, \, \, \, \, \, \, \, \, \, \, \, \, \, (r,u = 1,...,{\rm N})         
  \nonumber\\
  && \, \, \, \, \, \, \, \, \, \, \, \, \, \, \, \, \, 
  (l_{1} \hskip -0.015 truecm ,l_{2} = 0,1,... \, \, ; \, \, \, 
  m_{1 \hskip -0.015 truecm (2)} = -l_{1 \hskip -0.015 truecm (2)},...,l_{1 \hskip -0.015 truecm (2)}),     
\label{e29} 
\end{eqnarray}
which completes the proof. 

Substituting this equation into~(\ref{e23}) and using  
the facts that $\left \langle \, 0 \, 0 \, 0 \, 0 \, | \, 0 \, 0 \, \right \rangle = 1$, 
$Y_{0 0}({\Omega}_{{\vec R}_{sr}}) = 1/{\sqrt{4 \pi}}$, and 
$h^{(1)}_{0}(k R_{rs}) = -i \, \e^{ \, i k R_{rs}} \hskip -0.060 truecm /{k R_{rs}}$, we obtain the set of equations   
corresponding to the case considered, which is      
\begin{eqnarray}
  && {\cal T}_{0 0 \hskip +0.023 truecm ; \hskip +0.023 truecm 0 0}^{\, (r u)}(k \, ,k \, ;E)   
  + {(2 \pi)^{2}} \, \hskip -0.019 truecm \frac{m k}{{\hbar}^{2}} \, t_{0}(k \, ,k \, ;E) 
  \hskip -0.11 truecm
  \sum_{\scriptstyle s=1 \atop \scriptstyle (s \neq r)}^{\rm N} 
  \hskip -0.11 truecm
  \Biggl (\frac{\e^{ \, i k R_{rs}}}{k R_{rs}} \Biggr ) \, 
  {\cal T}_{0 0 \hskip +0.023 truecm ; \hskip +0.023 truecm 0 0}^{\, (s u)}(k \, ,k \, ;E)    
  \nonumber\\
  && = t_{0}(k \, ,k \, ;E) \, {\delta}_{r u}
  \nonumber\\  
  && \, \, \, \, \, \, \, \, \, \, \, \, \, \, \, \, \, \, \, \, \, \, \, \, \, \, \, \, \, \, \, \, \, \, \, \, \, 
  \, \, \, \, \, \, \, \, \, \, \, \, \, \, \, \, \, \, \, \, \, \, \, \, \, \, \, \, \, \, \, \, \, \, \, \, \, \, 
  \, \, \, \, \, \, \, \, \, \, \, \, \, \, \, \, \, \, \, \, \, \, \, \, \, \, \, \, \, \, \, \, \, \, \, \, \, \,
  \, \, \, \, \, \, \, \, \, \, \, \, \hskip -0.018 truecm (r,u = 1,...,{\rm N}).
\label{e30}
\end{eqnarray}

The function $t({\vec k}^{'} \hskip -0.040 truecm ,{\vec k} \, ;E)$ has no angular dependence if the scatterer is pointlike. A
pointlike scatterer can be modelled by any potential well of spherical symmetry whose
radius $a$ is arbitrarily small. Scattering by such a scatterer is characterized by the sole
phase shift ${\delta}_{0} = {\delta}$ because it is only for the wave function corresponding to $l = 0$ that 
the logarithmic derivative at $r = a$ has a limit when $a$ tends to $0$. The relation for the
dependence of ${\delta}$ on $k$ is obtained as follows. Since the wave function outside the well is
proportional to $\sin \hskip +0.030 truecm (kr + {\delta})$, its logarithmic derivative at $r = a$ tends to $k \cos \hskip -0.010 truecm {\delta} / \sin \hskip -0.039 truecm {\delta}$ in
the limit. Setting this ratio equal to the value $c$ to which the logarithmic derivative
of the wave function inside the well tends in the limit, one finds
\beq
  \cot \hskip -0.012 truecm {\delta} = \frac{c}{k}.        
\label{e31}
\eeq
The expression of the function $t({\vec k}^{'} \hskip -0.040 truecm ,{\vec k} \, ;E)$ for a pointlike scatterer is 
\begin{eqnarray}
  t({\vec k}^{'} \hskip -0.040 truecm ,{\vec k} \, ;E) = t_{0}(k \, ,k \, ;E)
  = - \frac{1}{(2 \pi)^{2}} \, \frac{{\hbar}^{2}}{m k} \sin \hskip -0.035 truecm {\delta} \, \e^{ \, i {\delta}}.
\label{e34}
\end{eqnarray}
Since this expression remains invariant if one replaces ${\delta}$ by ${\delta} + {\pi}$, the interval of variation 
of the phase shift can be restricted to $]0 \, , {\pi}[$. 

Substituting~(\ref{e34}) into~(\ref{e30}), we obtain   
\begin{eqnarray}
  && {\cal T}_{0 0 \hskip +0.023 truecm ; \hskip +0.023 truecm 0 0}^{\, (r u)}(k \, ,k \, ;E)    
  - \sin \hskip -0.035 truecm {\delta} \, \e^{ \, i {\delta}} 
  \hskip -0.11 truecm
  \sum_{\scriptstyle s = 1 \atop \scriptstyle (s \neq r)}^{\rm N}      
  \hskip -0.11 truecm
  \Biggl (\frac{\e^{ \, i k R_{rs}}}{k R_{rs}} \Biggr ) \, 
  {\cal T}_{0 0 \hskip +0.023 truecm ; \hskip +0.023 truecm 0 0}^{\, (s u)}(k \, ,k \, ;E) 
  \nonumber\\
  && = - \frac{1}{(2 \pi)^{2}} \, \frac{{\hbar}^{2}}{m k} \sin \hskip -0.035 truecm {\delta} \, \e^{ \, i {\delta}} \, {\delta}_{r u}
  \, \, \, \, \, \, \, \, \, \, \, \, \, \, \, \, \, \, \, \, \, \, \, \, \, \, \, \, \, \, \, 
  \, \, \, \, \, \, \, \, \, \, \, \, \, \, \, \, \, \, \, \, \, \, \, \, \, (r,u = 1,...,{\rm N}). 
\label{e35}
\end{eqnarray}
This set of ${\rm N}^{2}$ equations for the variables 
${\cal T}_{0 0 \hskip +0.023 truecm ; \hskip +0.023 truecm 0 0}^{\, (r u)}(k \, ,k \, ;E)$
can be reduced 
to a system of only ${\rm N}$ equations for the dimensionless variables ${\Theta}_{r}$ defined by  
\begin{eqnarray}
  && {\Theta}_{r} = - {(2 \pi)^{2}} \, \hskip -0.019 truecm \frac{m k}{{\hbar}^{2}} \, \hskip -0.030 truecm
  (\cot \hskip -0.012 truecm {\delta} - i)
  \sum_{u = 1}^{\rm N} \e^{ \, i {\vec k}.{\vec R}_{ru}} \,   
  {\cal T}_{0 0 \hskip +0.023 truecm ; \hskip +0.023 truecm 0 0}^{\, (r u)}(k \, ,k \, ;E) 
  \nonumber\\
  && \, \, \, \, \, \, \, \, \, \, \, \, \, \, \, \, \, \, \, \, \, \, \, \, \, \, \, \, \, \, \, \, \, \, \, \, 
  \, \, \, \, \, \, \, \, \, \, \, \, \, \, \, \, \, \, \, \, \, \, \, \, \, \, \, \, \, \, \, \, \, \, \, \, \,
  \, \, \, \, \, \, \, \, \, \, \, \, \, \, \, \, \, \, \, \, \, \, \, \, \, \, \, \, \, \, \, \, \, \, \, \, \,
  \, \, \, \, \, \, \, \, \, \, \, \, \, \, \, \,
  \, (r = 1,...,{\rm N}).
\label{e36}
\end{eqnarray}
The proof is as follows. Multiplying each equation by the factor 
$\e^{ \, i {\vec k}.{\vec R}_{ru}} = \e^{ \, i {\vec k}.{\vec R}_{rs}} \, \e^{ \, i {\vec k}.{\vec R}_{su}}$ 
and then adding the equations with the same value of the index $r$, we obtain the set of ${\rm N}$ equations  
\begin{eqnarray}
  && \sum_{u = 1}^{\rm N} \e^{ \, i {\vec k}.{\vec R}_{ru}} \,  
  {\cal T}_{0 0 \hskip +0.023 truecm ; \hskip +0.023 truecm 0 0}^{\, (r u)}(k \, ,k \, ;E)     
  \nonumber\\  
  && \, \, \, - \sin \hskip -0.035 truecm {\delta} \, \e^{ \, i {\delta}} 
  \hskip -0.11 truecm
  \sum_{\scriptstyle s = 1 \atop \scriptstyle (s \neq r)}^{\rm N}      
  \hskip -0.11 truecm
  \Biggl (\frac{\e^{ \, i k R_{rs}}}{k R_{rs}} \Biggr ) \, \e^{ \, i {\vec k}.{\vec R}_{rs}}
  \sum_{u = 1}^{\rm N} \e^{ \, i {\vec k}.{\vec R}_{su}} \, 
  {\cal T}_{0 0 \hskip +0.023 truecm ; \hskip +0.023 truecm 0 0}^{\, (s u)}(k \, ,k \, ;E)  
  \nonumber\\
  && =  - \frac{1}{(2 \pi)^{2}} \, \frac{{\hbar}^{2}}{m k} \sin \hskip -0.035 truecm {\delta} \, \e^{ \, i {\delta}} 
  \, \, \, \, \, \, \, \, \, \, \, \, \, \, \, \, \, \, \, \, \, \, \, \, \, \, \, \, \, \, \, \, \, \, \,
  \, \, \, \, \, \, \, \, \, \, \, \, \, \, \, \, \, \, \, \, \, \, \, \, \, \, \, \, \, \, \, \, \, \, \, (r = 1,...,{\rm N}).
\label{e37}
\end{eqnarray}
Each of these equations can be expressed in terms of the new variables ${\Theta}_{r}$. This
leads to the system of equations
\begin{eqnarray}
  && {\Theta}_{r} - \sin \hskip -0.035 truecm {\delta} \, \e^{ \, i {\delta}} 
  \hskip -0.11 truecm
  \sum_{\scriptstyle s = 1 \atop \scriptstyle (s \neq r)}^{\rm N}
  \hskip -0.11 truecm
  \Biggl (\frac{\e^{ \, i k R_{rs}}}{k R_{rs}} \Biggr ) \, \e^{ \, i {\vec k}.{\vec R}_{rs}} \, {\Theta}_{s} = 1
  \nonumber\\
  && \, \, \, \, \, \, \, \, \, \, \, \, \, \, \, \, \, \, \, \, \, \, \, \, \, \, \, \, \, \, \, \, \, \, \, \, 
  \, \, \, \, \, \, \, \, \, \, \, \, \, \, \, \, \, \, \, \, \, \, \, \, \, \, \, \, \, \, \, \, \, \, \, \, \,
  \, \, \, \, \, \, \, \, \, \, \, \, \, \, \, \, \, \, \, \, \, \, \, \, \, \, \, \, \, \, \, \, \, \, \, \, \,
  \, \, \, \, \, \, \, \, \, \, \, \, \, \, \, \,  
  \, \, (r = 1,...,{\rm N}),
\label{e38}
\end{eqnarray}
which completes the proof. The reduced set of equations is obviously better suited for calculations
than the original one.

Since the functions ${\cal T}_{ru}({\vec k}^{'} \hskip -0.040 truecm ,{\vec k} \, ;E)$ have no angular 
dependence, the expression of the scattering amplitude is very simple for pointlike scatterers. Substituting  
${\cal T}_{0 0 \hskip +0.023 truecm ; \hskip +0.023 truecm 0 0}^{\, (r u)}(k \, ,k \, ;E)$ for 
${\cal T}_{ru}({\vec k}^{'} \hskip -0.040 truecm ,{\vec k} \, ;E)$ in~(\ref{e12}) and using the 
definition of the variables ${\Theta}_{r}$, we find   
\beq
  f({\vec k}^{'} \hskip -0.040 truecm ;{\vec k}) = \frac{1}{k} \sin \hskip -0.035 truecm {\delta} \, \e^{ \, i {\delta}} 
  \sum_{r=1}^{\rm N} \e^{ \, -i ({\vec k}^{'} \hskip -0.095 truecm -{\vec k}).{\vec R}_{r}} \, {\Theta}_{r}. 
\label{e39}
\eeq

\section{Cross section}

The cross section ${\sigma}_{\rm tot}$ for scattering by the target (the subscript is an abbreviation for
"tot(ality of the scatterers)") is obtained by integrating $\abs{f({\vec k}^{'} \hskip -0.040 truecm ;{\vec k})}^{2}$
over ${\Omega}_{\vec k^{'}}$. We find                                                       
\begin{eqnarray}
  {\sigma}_{\rm tot} = {\sigma} \Biggl \lbrack \, \sum_{r = 1}^{\rm N} \abs{{\Theta}_{r}}^{2}  
  + \sum_{r = 1}^{\rm N} 
  \hskip -0.11 truecm
  \sum_{\scriptstyle s = 1 \atop \scriptstyle (s \neq r)}^{\rm N} 
  \hskip -0.11 truecm
  \Biggl (\frac{\sin \hskip +0.010 truecm (k R_{rs})}{k R_{rs}} \Biggr ) \,
  \e^{ \, i {\vec k}.{\vec R}_{rs}} \, {\Theta}_{r}^{*} \, {\Theta}_{s} \Biggr \rbrack,
\label{e40}   
\end{eqnarray}
where ${\sigma} = (4 {\pi}/k^{2}) \sin^{2} \hskip -0.045 truecm {\delta}$ is the cross section for scattering 
by an individual scatterer. 
   
The optical theorem~\cite{s} is not only verified for a single pointlike scatterer, as one
has $(k/{4 \pi}) \hskip +0.050 truecm {\sigma} =
\Im \lbrace (1/k) \sin \hskip -0.035 truecm {\delta} \, \e^{ \, i {\delta}} \rbrace$, but also for a set of any number
of them. The proof 
is as follows. Multiplication of each equation in~(\ref{e38}) by ${\Theta}_{r}^{*}$ gives 
\begin{eqnarray}
  && \abs{{\Theta}_{r}}^{2} - \sin \hskip -0.035 truecm {\delta} \, \e^{ \, i {\delta}} 
  \hskip -0.11 truecm
  \sum_{\scriptstyle s = 1 \atop \scriptstyle (s \neq r)}^{\rm N}
  \hskip -0.11 truecm
  \Biggl (\frac{\e^{ \, i k R_{rs}}}{k R_{rs}} \Biggr ) \, \e^{ \, i {\vec k}.{\vec R}_{rs}} \, {\Theta}_{r}^{*} \,    
  {\Theta}_{s} = {\Theta}_{r}^{*} 
  \nonumber\\  
  && \, \, \, \, \, \, \, \, \, \, \, \, \, \, \, \, \, \, \, \, \, \, \, \, \, \, \, \, \, \, \, \, \, \, \, \, 
  \, \, \, \, \, \, \, \, \, \, \, \, \, \, \, \, \, \, \, \, \, \, \, \, \, \, \, \, \, \, \, \, \, \, \, \, \,
  \, \, \, \, \, \, \, \, \, \, \, \, \, \, \, \, \, \, \, \, \, \, \, \, \, \, \, \, \, \, \, \, \, \, \, \, \, 
  \, \, \, \, \, \, \, \, \, \, \, \, \, \, \, \, \, \, \, \, \,
  (r = 1,...,{\rm N}).
\label{e41}
\end{eqnarray}
Adding all these equations and taking the real and the imaginary part of the resulting equality, we obtain
the two equalities    
\begin{eqnarray}
  \sum_{r = 1}^{\rm N} \abs{{\Theta}_{r}}^{2} 
  - \sin \hskip -0.035 truecm {\delta} \sum_{r = 1}^{\rm N} 
  \hskip -0.11 truecm
  \sum_{\scriptstyle s = 1 \atop \scriptstyle (s \neq r)}^{\rm N}
  \hskip -0.11 truecm
  \Biggl (\frac{\cos \hskip +0.0327 truecm (k R_{rs} + {\delta})}{k R_{rs}} \Biggr ) \, \e^{ \, i {\vec k}.{\vec R}_{rs}} \, 
  {\Theta}_{r}^{*} \, {\Theta}_{s} = \sum_{r = 1}^{\rm N} \Re \lbrace {\Theta}_{r} \rbrace
  \nonumber\\
\label{e42}
\end{eqnarray}
and
\begin{eqnarray} 
  && \sin \hskip -0.035 truecm {\delta} \sum_{r = 1}^{\rm N} 
  \hskip -0.11 truecm
  \sum_{\scriptstyle s = 1 \atop \scriptstyle (s \neq r)}^{\rm N}
  \hskip -0.11 truecm
  \Biggl (\frac{\sin \hskip +0.010 truecm (k R_{rs} + {\delta})}{k R_{rs}} \Biggr ) \, \e^{ \, i {\vec k}.{\vec R}_{rs}} \, 
  {\Theta}_{r}^{*} \, {\Theta}_{s} = \sum_{r = 1}^{\rm N} \Im \lbrace {\Theta}_{r} \rbrace. 
\label{e43}
\end{eqnarray}
Multiplying the first and the second equality by $\sin \hskip -0.035 truecm {\delta}$ 
and $\cos \hskip -0.0129 truecm {\delta}$ respectively and adding, we find 
\begin{eqnarray}
  && \sin \hskip -0.035 truecm {\delta} \sum_{r = 1}^{\rm N} \abs{{\Theta}_{r}}^{2}  
  + \sin \hskip -0.035 truecm {\delta} \sum_{r = 1}^{\rm N} 
  \hskip -0.11 truecm
  \sum_{\scriptstyle s = 1 \atop \scriptstyle (s \neq r)}^{\rm N} 
  \hskip -0.11 truecm
  \Biggl (\frac{\sin \hskip +0.010 truecm (k R_{rs})}{k R_{rs}} \Biggr ) \, \e^{ \, i {\vec k}.{\vec R}_{rs}} \, 
  {\Theta}_{r}^{*} \, {\Theta}_{s}   
  \nonumber\\
  && = \sin \hskip -0.035 truecm {\delta} \sum_{r = 1}^{\rm N} \Re \lbrace {\Theta}_{r} \rbrace
  + \cos \hskip -0.0129 truecm {\delta} \sum_{r = 1}^{\rm N} \Im \lbrace {\Theta}_{r} \rbrace.
\label{e44}
\end{eqnarray}   
Substitution of this new equality into~(\ref{e40}) leads to
\begin{eqnarray}  
  \frac{k}{4 \pi} \, \hskip -0.020 truecm {\sigma}_{\rm tot} && = \frac{1}{k} \sin \hskip -0.035 truecm {\delta}
  \Biggl \lbrack \hskip +0.024 truecm \sin \hskip -0.035 truecm {\delta} \sum_{r = 1}^{\rm N} \Re \lbrace {\Theta}_{r} \rbrace 
  + \cos \hskip -0.0129 truecm {\delta} \sum_{r = 1}^{\rm N} \Im \lbrace {\Theta}_{r} \rbrace \hskip -0.015 truecm
  \Biggr \rbrack  
  \nonumber\\
  && = \Im \lbrace f({\vec k} \, ;{\vec k}) \rbrace,
\label{e45}
\end{eqnarray}      
which completes the proof.  
     
The cross section can also be written as a sum of terms each of which but the last gives the contribution  
of an increasing number ${\rm m}$ of successive scatterings. This sum is obtained by repeated
substitution of the system of equations into~(\ref{e45}). We find   
\begin{eqnarray}  
  {\sigma}_{\rm tot} && = {\rm N} {\sigma} 
  \nonumber\\ 
  && \, \, \, + \frac{4 \pi}{k^{2}} \sum_{{\rm m} = 2}^{\rm M} \, (\sin \hskip -0.035 truecm {\delta})^{\rm m}
  \hskip +0.010 truecm \Im \Biggl \lbrace \e^{ \, i {\rm m} {\delta}}
  \sum_{r_{1} \hskip -0.015 truecm = 1}^{\rm N} 
  \hskip -0.11 truecm
  \sum_{\scriptstyle r_{2} = 1 \atop \scriptstyle (r_{2} \neq r_{1} \hskip -0.015 truecm )}^{\rm N} \dots  
  \sum_{\scriptstyle r_{\rm m} = 1 \atop \scriptstyle (r_{\rm m} \neq r_{{\rm m}-1} \hskip -0.015 truecm )}^{\rm N}
  \hskip -0.11 truecm
  \Biggl (\frac{\e^{ \, i k R_{r_{1} \hskip -0.025 truecm r_{2}}}}{k R_{r_{1} \hskip -0.025 truecm r_{2}}} \Biggr )
  \nonumber\\
  && \, \, \, \, \, \, \, \,              
  \times \Biggl (\frac{\e^{ \, i k R_{r_{2}r_{3}}}}{k R_{r_{2}r_{3}}} \Biggr )  
  \dots \Biggl (\frac{\e^{ \, i k R_{r_{{\rm m}-1} \hskip -0.025 truecm r_{\rm m}}}}
  {k R_{r_{{\rm m}-1} \hskip -0.025 truecm r_{\rm m}}} \Biggr ) \,  
  \e^{ \, i{\vec k}.{\vec R}_{r_{1} \hskip -0.025 truecm r_{\rm m}}} \Biggr \rbrace 
  \nonumber\\
  && \, \, \, + \frac{4 \pi}{k^{2}} \, (\sin \hskip -0.035 truecm {\delta})^{{\rm M}+1} \hskip -0.002 truecm
  \Im \Biggl \lbrace \e^{ \, i ({\rm M}+1) {\delta}}
  \sum_{r_{1} \hskip -0.015 truecm = 1}^{\rm N} 
  \hskip -0.11 truecm
  \sum_{\scriptstyle r_{2} = 1 \atop \scriptstyle (r_{2} \neq r_{1} \hskip -0.015 truecm )}^{\rm N} \dots 
  \sum_{\scriptstyle r_{{\rm M}+1} \hskip -0.015 truecm = 1 
  \atop \scriptstyle (r_{{\rm M}+1} \hskip -0.015 truecm \neq r_{\rm M})}^{\rm N}
  \hskip -0.11 truecm
  \Biggl (\frac{\e^{ \, i k R_{r_{1} \hskip -0.025 truecm r_{2}}}}{k R_{r_{1} \hskip -0.025 truecm r_{2}}} \Biggr )  
  \nonumber\\  
  && \, \, \, \, \, \, \, \,            
  \times \Biggl (\frac{\e^{ \, i k R_{r_{2}r_{3}}}}{k R_{r_{2}r_{3}}} \Biggr )
  \dots \Biggl (\frac{\e^{ \, i k R_{r_{\rm M}r_{{\rm M}+1}}}}{k R_{r_{\rm M}r_{{\rm M}+1}}} \Biggr ) \,  
  \e^{ \, i{\vec k}.{\vec R}_{r_{1} \hskip -0.025 truecm r_{{\rm M}+1}}} \, {\Theta}_{r_{{\rm M}+1}} \Biggr \rbrace,   
\label{e46}
\end{eqnarray} 
where ${\rm M}$ is arbitrary. The contribution of a given number ${\rm m}$ of scatterings 
to the cross section consists of the sum of those of ${\rm N}({\rm N}-1)^{{\rm m}-1}$ different sequences of  
scatterings. The first term in the expression provides the contribution of a single scattering, which is given 
by the sum of the individual cross sections as the particle can be scattered by any of the scatterers. The last 
term gives the expression of the difference between the exact expression of the cross section and the sum 
of contributions corresponding to ${\rm M}$ substitutions, and so that of the remainder in the approximation 
of the former by the latter. In the simplest approximation the expression of the cross section reduces to 
the contribution of a single scattering; this approximation is obviously more accurate the larger
the distance between the closest scatterers.

\section{Numerical study}

The fact that the set of equations to be solved in order to obtain one value of the cross
section is finite when the scatterers are pointlike makes this type of scatterer especially 
suitable for numerical studies in which a large number of such values is needed. In this 
section we present and discuss the results of a numerical study about the scattering by
a disordered target consisting of pointlike scatterers. The target has been modelled by
a set of ${\rm N}$ scatterers having each an equal probability of being at any position inside a
sphere whose radius ${\rm R}$ may be modified; samples fitting in the same volume have been 
assigned the same value of the density ${\rho} = 3 {\rm N}/4 {\pi} {\rm R}^{3}$. We have studied the 
variation of
the normalized mean cross 
section ${\left \langle {\sigma}_{\rm tot} \right \rangle}/ \hskip +0.030 truecm {\rm N} {\sigma}$ with the 
density ${\rho}$ of the samples, and so
with the radius ${\rm R}$ of the target, for different values of the phase shift. The value of the 
cross section for a sample at a particular value of the phase shift has been obtained by
substituting the solution of~(\ref{e38}) into~(\ref{e45}), which gives an expression of ${\sigma}_{\rm tot}$ linear in
the variables. The value of the mean cross section corresponding to a particular value
of the density and of the phase shift has been obtained by calculating the mean of the   
values of the cross section of $300$ different samples with these values of the density and
the phase shift. The calculations have been done for ${\rm N} = 100$ and with $m = k = {\hbar} = 1$. 
The values of the phase shift have been taken in the particular interval $[{\pi}/2 \, , {\pi}[$, the reason being that 
the features of the mean cross section discussed in this paper do not only exist for any
value in this interval but also for any value in $]0 \, , {\pi}/2[$; for values of the phase shift that belong to $]0 \, , {\pi}/2[$, the 
mean cross section has additional features, which will be discussed in a separate paper. 
The results of the numerical study are shown in the two figures below.   
   
Figures 1 and 2 show the variation of the normalized mean cross section with the density
of the samples for two sets of values of the phase shift (the scale on both axes is 
logarithmic in each figure). 

\begin{figure}[!htbp] 
\begin{center}
\includegraphics[angle=270,width=.67\linewidth,clip=true]{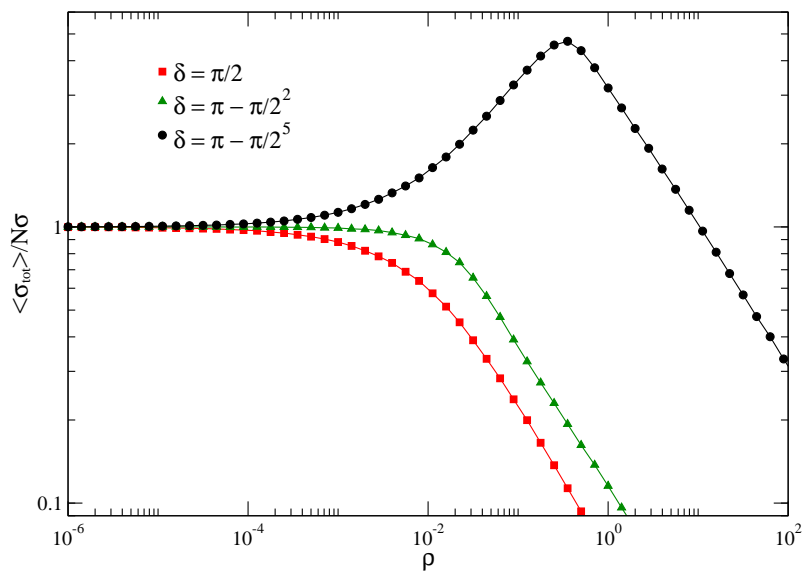}
\caption{\small
Variation of the normalized mean cross section 
${\left \langle {\sigma}_{\rm tot} \right \rangle}/ \hskip +0.030 truecm {\rm N} {\sigma}$
with the density ${\rho}$ of the samples for a set  
of values of the phase shift.}  
\label{fig1}
\end{center}
\end{figure}

\begin{figure}[!htbp] 
\begin{center}
\includegraphics[angle=270,width=.67\linewidth,clip=true]{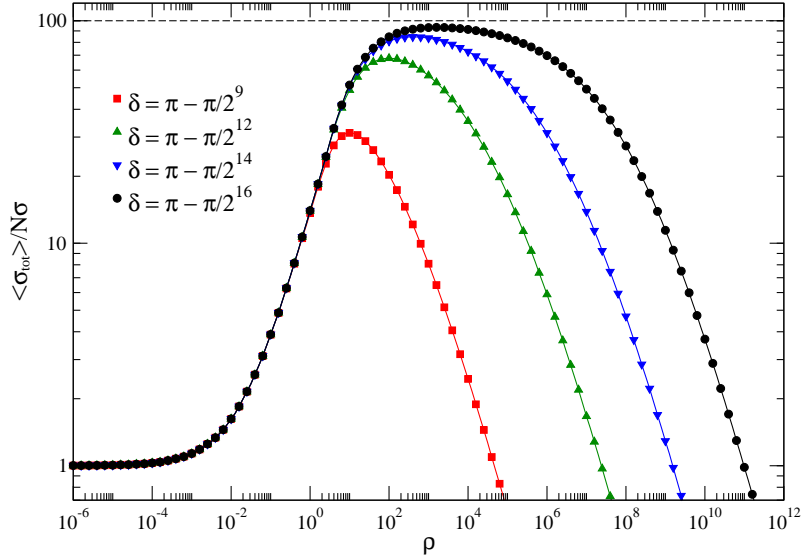}
\caption{\small
Variation of the normalized mean cross section 
${\left \langle {\sigma}_{\rm tot} \right \rangle}/ \hskip +0.030 truecm {\rm N} {\sigma}$
with the density ${\rho}$ of the samples for another set  
of values of the phase shift. The chosen number of scatterers is shown by an horizontal dashed line.}   
\label{fig2}
\end{center}
\end{figure} 

\noindent The obtained curves have a common feature, namely that the value of the normalized 
mean cross section is close to $1$ when the density is very small, and so the radius very 
large. This is a consequence of the fact that the distance between the closest scatterers
is sufficiently large in a typical arrangement of very low density for the contribution of 
a single scattering to be the largest one and so for the cross section to be almost equal 
to ${\rm N}{\sigma}$. The figures show also that the mean cross section first either decreases (Fig. 1)
or increases (in the two figures) as the density increases; moreover, it always decreases 
uniformly as the density increases further on. The detailed explanation for each one of 
these two, less expected features of the mean cross section is given in the remainder of
the section.
 
The decrease and the increase are both originated by double scattering; the reason
for which the mean cross section can first deviate from ${\rm N}{\sigma}$ in opposite ways is provided
by its expression in the approximation of single and double scattering. This expression 
is derived in Appendix A. The result is 
\begin{eqnarray}    
  {\left \langle {\sigma}_{{\rm s} + {\rm d}} \right \rangle} = {\rm N} {\sigma}    
  + \frac{9 {\rm N} ({\rm N} -1)}{4(k {\rm R})^{6}} \biggl \lbrack \hskip +0.024 truecm
  \sin (2 {\delta}) \hskip +0.060 truecm
  {\rm C}(k {\rm R}) 
  + \cos \hskip +0.022 truecm (2 {\delta}) \, {\rm S}(k {\rm R}) \hskip -0.011 truecm \biggr \rbrack {\sigma},  
\label{e49}
\end{eqnarray}
(the subscript "tot" has been replaced by "s+d", which is an abbreviation for "s(ingle)
and d(ouble scattering)", in order to avoid any confusion with the exact cross section),
with
\beq
  {\rm C}(k {\rm R}) = \frac{(k {\rm R}) \cos \hskip +0.022 truecm (4 k {\rm R})}{16} - \frac{\sin (4 k {\rm R})}{64} 
  + \frac{(k {\rm R})^{3}}{3} 
\label{e50}
\eeq
and
\beq 
  {\rm S}(k {\rm R}) = \frac{(k {\rm R}) \sin (4 k {\rm R})}{16} + \frac{\cos \hskip +0.022 truecm (4 k {\rm R})}{64} 
  + \frac{(k {\rm R})^{4}}{2} - \frac{(k {\rm R})^{2}}{8} - \frac{1}{64}.
\label{e51}
\eeq 
\noindent When the radius is large enough, and so the density small enough, the contribution of 
double scattering is dominated by its term of least power, which is the one in $1/(k {\rm R})^{2}$.
The normalized mean cross section is then well approximated by
\begin{eqnarray}    
  {\left \langle {\sigma}_{\rm tot} \right \rangle} / {\rm N} {\sigma} && = 1
  + \frac{9}{8} ({\rm N} -1) \, \frac{\cos \hskip +0.022 truecm (2 {\delta})}{(k {\rm R})^{2}} 
  \nonumber\\ 
  && = 1 + \frac{1}{2} ({\rm N} -1) \cos \hskip +0.022 truecm (2 {\delta}) 
  \biggl (\frac{9 {\pi} {\rho}}{2 {\rm N} k^{3}} \biggr )^{ \hskip -0.076 truecm \frac{2}{3}}.   
\label{e52}
\end{eqnarray} 
This equation shows that if the phase shift takes on its values in $]{\pi}/4 \, , 3{\pi}/4[$, 
the contribution of double scattering is less than $0$, which implies that the mean cross
section decreases as the density increases (Fig. 1); on the contrary, if the phase shift takes on its values 
either in $]0 \, , {\pi}/4[$ or $]3{\pi}/4 \, , {\pi}[$, this contribution is greater than $0$, with the consequence that the mean cross section increases as the density 
increases (in the two figures).   

The fact that both the increase and the decrease are caused by double scattering
alone is illustrated by the next two figures, in which the variation of  ${\left \langle {\sigma}_{\rm tot} \right \rangle}$ is compared 
to that of ${\left \langle {\sigma}_{{\rm s} + {\rm d}} \right \rangle}$ for two different values of the phase shift, one belonging 
to the interval $]{\pi}/4 \, , 3{\pi}/4[$ (Fig. 3) and the other to the interval $]3{\pi}/4 \, , {\pi}[$ (Fig. 4).
One sees that ${\left \langle {\sigma}_{\rm tot} \right \rangle}$ has the same behaviour as
${\left \langle {\sigma}_{{\rm s} + {\rm d}} \right \rangle}$, deviating from ${\rm N}$ by decreasing
when ${\delta}$ belongs to $]{\pi}/4 \, , 3{\pi}/4[$ and increasing when it belongs to $]3{\pi}/4 \, , {\pi}[$, 
in agreement with~(\ref{e52}). In addition,
Fig. 4 shows that the mean cross section is not bounded in the approximation of single 
and double scattering, which comes from the fact that the ratio ${\rm C}(k {\rm R})/(k {\rm R})^{6}$ diverges  
when the radius becomes arbitrarily small.

\begin{figure}[!htbp] 
\begin{center}
\includegraphics[angle=270,width=.67\linewidth,clip=true]{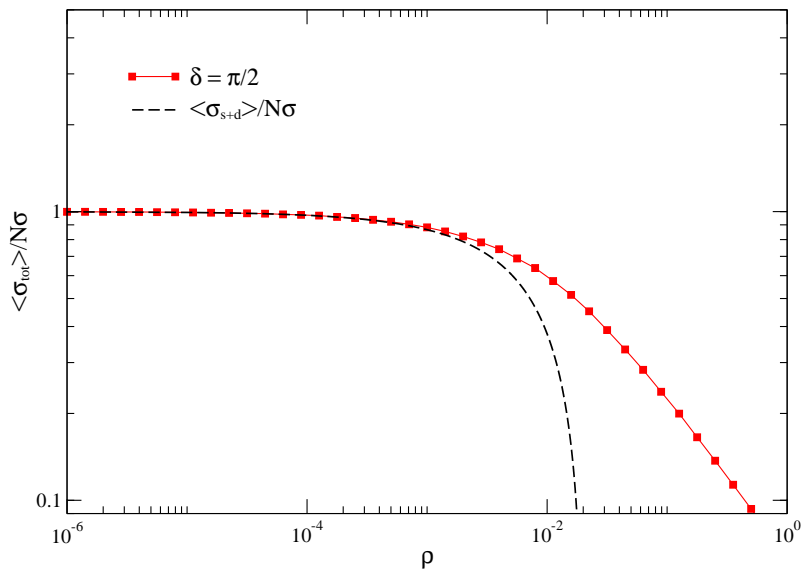}
\caption{\small
Comparison of the variations of ${\left \langle {\sigma}_{\rm tot} \right \rangle}/{{\rm N} {\sigma}}$ and
${\left \langle {\sigma}_{{\rm s} + {\rm d}} \right \rangle}/{{\rm N} {\sigma}}$ 
with the density ${\rho}$ of the samples for a value of the phase shift belonging to the interval $]{\pi}/4 \, , 3{\pi}/4[$.}
\label{fig3}
\end{center}
\end{figure}

\begin{figure}[!htbp] 
\begin{center}
\includegraphics[angle=270,width=.67\linewidth,clip=true]{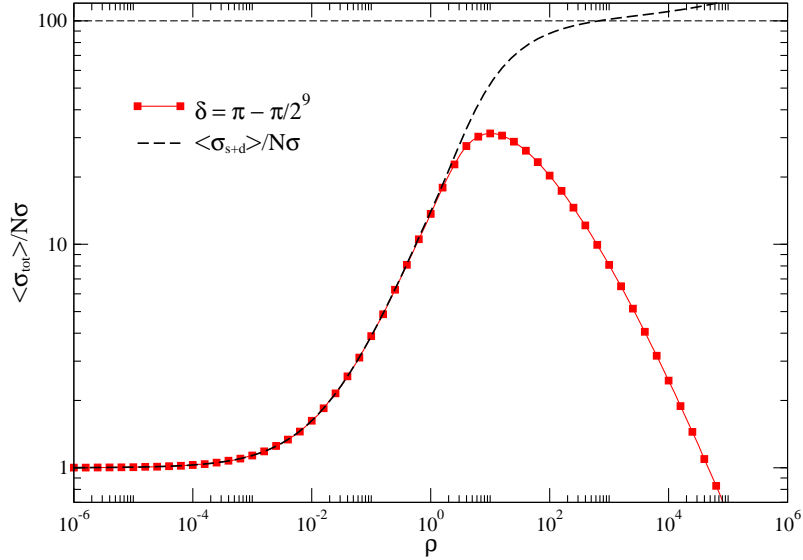}
\caption{\small
Comparison of the variations of ${\left \langle {\sigma}_{\rm tot} \right \rangle}/{{\rm N} {\sigma}}$ and
${\left \langle {\sigma}_{{\rm s} + {\rm d}} \right \rangle}/{{\rm N} {\sigma}}$ 
with the density ${\rho}$ of the samples for a value of the phase shift belonging to the interval $]3{\pi}/4 \, , {\pi}[$.} 
\label{fig4}
\end{center}
\end{figure}

The uniform decrease of the mean cross section is caused by an unlimited number of contributions of successive 
scatterings to the cross section, that is to say, by multiple scattering. The expression of the mean cross section  
corresponding to an unlimited number of contributions of successive scatterings is derived in detail in Appendix B.
We~find
\begin{eqnarray}
  && \Biggl (\frac{k^{2}}{4 \pi} \Biggr ) \Biggl (\frac{{{\rm N}-1}}{{\rm N}} \Biggr )
  {\left \langle {\sigma}_{\rm tot} \right \rangle} 
  \nonumber\\
  && = \sum_{l=0}^{\infty} (2l+1) \, 
  \Im \Biggl \lbrace i \Biggl \lbrack \frac{({\cal K} {\rm R}) \, j_{l}(k{\rm R}) \, j_{l+1}({\cal K} {\rm R})
  - (k {\rm R}) \, j_{l}({\cal K} {\rm R}) \, j_{l+1}(k{\rm R})}
  {({\cal K} {\rm R}) \,  h_{l}^{(1)}(k{\rm R}) \, j_{l+1}({\cal K} {\rm R}) - (k {\rm R}) \, j_{l}({\cal K} {\rm R}) \, h_{l+1}^{(1)}(k{\rm R})}
  \Biggr \rbrack \Biggr \rbrace,      
  \nonumber\\
  \nonumber\\ 
\label{e53}
\end{eqnarray}
with 
\beq
  {\cal K} = k \, \sqrt {1 + \frac{3 ( {\rm N}-1 ) \sin \hskip -0.035 truecm {\delta} \, \e^{ \, i
  {\delta}}}{(k{\rm R})^{3}}}.
\label{e54}
\eeq

In Figures 5 and 6 the mean cross section is compared to the curve obtained from the expression in~(\ref{e53}),
in which the sum has been truncated at $l = 150$,
for the values of the phase shift that have been taken in the two previous figures.

\begin{figure}[!tbp] 
\begin{center}
\includegraphics[angle=270,width=.67\linewidth,clip=true]{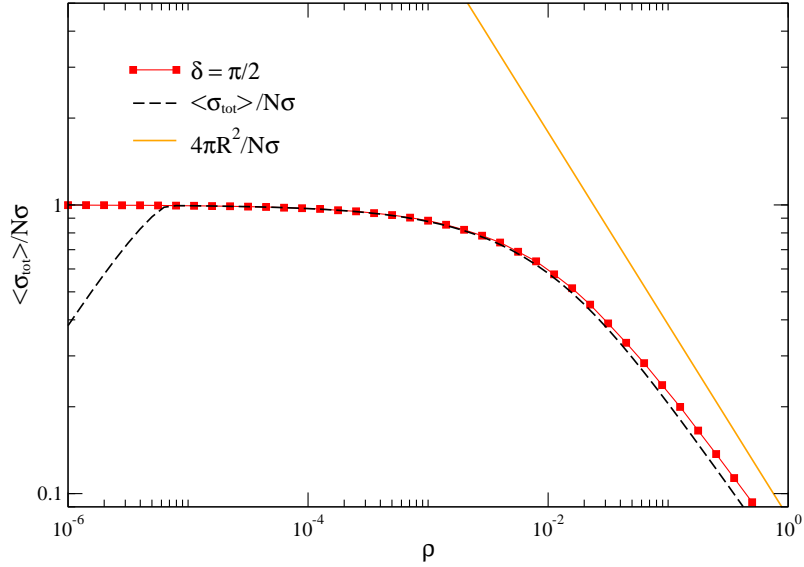}
\caption{\small
Comparison of the variations of the normalized mean cross section and of the expression of ${\left \langle {\sigma}_{\rm tot} \right \rangle}/{{\rm N} {\sigma}}$
given by~(\ref{e53}), in which the sum has been truncated at $l = 150$ (dashed line), for a value of the phase shift belonging to the interval $]{\pi}/4 \, , 3{\pi}/4[$. The 
full line gives the values of the normalized cross section for diffusion by a hard sphere of radius ${\rm R}$.}  
\label{fig5}
\end{center}
\end{figure}

Each figure shows that the expression in~(\ref{e53}) reproduces well the mean cross section
except in the uninteresting range of values of the density in which the approximation of single scattering is valid. This 
discrepancy reflects the fact that the contributions of more and more partial waves have to be taken into account 
in the sum as the radius increases, and so as the density decreases. One can therefore conclude that the  
uniform decrease is caused by multiple scattering.

It is interesting to derive the expression to which~(\ref{e53}) reduces
when ${\rm R}$ tends to 0. We find
\begin{eqnarray}  
  {\left \langle {\sigma}_{\rm tot} \right \rangle} \approx \frac{4 \pi}{k^{2}} 
  \Im \biggl \lbrace j_{0}(k {\rm R}) \, (k {\rm R}) \, \e^{ \, i k {\rm R}} \biggr \rbrace  
  \approx 4 \pi {\rm R}^{2}.
\label{e55}
\end{eqnarray}  
This means that when the value of the density becomes arbitrarily large, the mean cross section 
tends towards the cross section for diffusion by a hard sphere whose radius is the same as that of  
the target, a behaviour which is clearly observed in Figures 5 and 6.

\begin{figure}[!tbp] 
\begin{center}
\includegraphics[angle=270,width=.67\linewidth,clip=true]{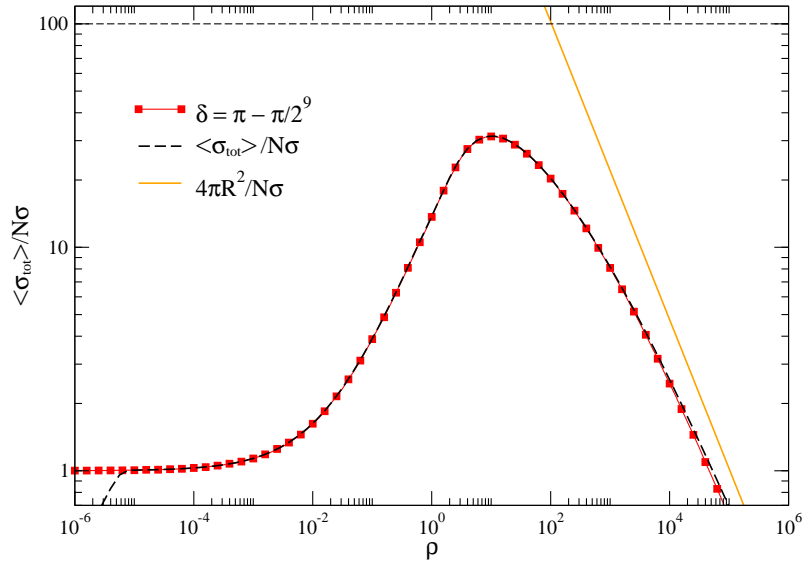}
\caption{\small
Comparison of the variations of the normalized mean cross section and of the expression of ${\left \langle {\sigma}_{\rm tot} \right \rangle}/{{\rm N} {\sigma}}$
given by~(\ref{e53}), in which the sum has been truncated at $l = 150$ (dashed line), for a value of the phase shift belonging to the interval $]3{\pi}/4 \, , {\pi}[$. The 
full line gives the values of the normalized cross section for diffusion by a hard sphere of radius ${\rm R}$.} 
\label{fig6}
\end{center}
\end{figure}

\section{Summary}

In this paper, we have developed a formalism which describes the quantum scattering of a particle 
by a disordered target consisting of pointlike scatterers.
The formalism has the important feature of preserving unitary because it 
takes into account all sequences of multiple scattering.
We have used it in a numerical study about the scattering by
a disordered target modelled by a set of pointlike scatterers which have each an equal probability
of being at any position inside a sphere whose radius may be modified.
We have studied the variation of the mean cross section with the          
density of the samples, and so with the radius of the target, for different values of the phase shift.
The study has shown that the mean cross section goes through successive stages as the density increases, and so 
as the radius decreases. The mean cross section is nearly constant at very low density, which reflects 
the fact that the particle is only scattered once before leaving the target.  
Depending on the value of the phase shift, the mean cross section either increases or decreases as the density increases, 
each one of the two behaviours being originated 
by double scattering. As the density increases further on, the mean cross section decreases uniformly whatever the value of the phase shift, 
a behaviour which is caused by multiple scattering and which follows that of the cross section for 
diffusion by a hard sphere potential of decreasing radius. 

\appendix
\setcounter{equation}{0}
\def\theequation{A.\arabic{equation}}
\def\thesubsection{A.\arabic{subsection}}
\section*{Appendix A: Expression of the mean cross section in the approximation of single and double scattering}

This appendix is devoted to the derivation of the expression of the mean cross section 
that includes only the contributions of single and double ({\rm m} = $2$) scattering. It follows 
from~(\ref{e46}) that the mean cross section is given in this particular approximation by
\begin{eqnarray}  
  {\left \langle {\sigma}_{{\rm s} + {\rm d}} \right \rangle} = {\rm N} {\sigma} 
  + {\sigma} \hskip -0.015 truecm \Im \Biggl \lbrace \e^{ \, 2 i {\delta}}
  \sum_{r = 1}^{\rm N}  
  \hskip -0.11 truecm  
  \sum_{\scriptstyle s = 1 \atop \scriptstyle (s \neq r)}^{\rm N}  
  \hskip -0.11 truecm   
  {\left \langle \Biggl (\frac{\e^{ \, i k R_{rs}}}{k R_{rs}} \Biggr ) \,
  \e^{ \, i{\vec k}.{\vec R}_{rs}} \right \rangle} \Biggr \rbrace    
\label{a1}
\end{eqnarray}   
(the subscript is for "s(ingle) and d(ouble scattering)"); finding its expression amounts
therefore to hardly more than finding that of the mean  
of ${({\e^{ \, i k R_{rs}}}/{k R_{rs}}) \, \e^{ \, i{\vec k}.{\vec R}_{rs}}}$.

Since each scatterer has an equal probability of being at any position in the target
(whose center is taken as the origin of coordinates), we have
\begin{eqnarray}
  {\left \langle \Biggl (\frac{\e^{ \, i k R_{rs}}}{k R_{rs}} \Biggr ) \,
  \e^{ \, i{\vec k}.{\vec R}_{rs}} \right \rangle} 
  && = \frac{1}{{\rm V}^{2}} \int_{\rm V} {\d {\vec R}_{r}} \int_{\rm V} {\d {\vec R}_{s}} \,
  \frac{\e^{ \, i k R_{rs}} \, \e^{ \, i {\vec k}.{\vec R}_{rs}}}{k R_{rs}}
  \nonumber\\
  && = \frac{1}{4 \pi} \, \frac{1}{4 \pi} \, \frac{9}{{\rm R}^{6}} \int_{\rm V} {\d {\vec R}_{r}} \int_{\rm V} {\d {\vec R}_{s}} \,
  \frac{\e^{ \, i k R_{rs}} \, \e^{ \, i {\vec k}.{\vec R}_{rs}}}{k R_{rs}}.   
\label{a2}
\end{eqnarray}
where ${\rm V}$ and ${\rm R}$ are the volume and radius of the target. The integration is most easily
done working with the set of angular variables that consists of the azimuthal and polar 
angles ${\Theta}_{rs}$ 
and ${\Phi}_{rs}$ of the vector ${\vec R}_{rs}$, the angle ${\theta}_{rs}$
between the vectors ${\vec R}_{r}$ and ${\vec R}_{s}$, and
the angle ${\phi}_{rs}$ between the planes that are defined by the 
vectors ${\vec R}_{r}$ and ${\vec R}_{s}$ and by the
vectors ${\vec R}_{rs}$ and ${\vec k}$ (which is assumed to determine the direction of the 
third axis). The
expression of the product of volume elements for this choice of angular variables is
\beq
  {\d {\vec R}_{r}} \hskip +0.040 truecm {\d {\vec R}_{s}} = {R^{2}_{r}} \hskip +0.040 truecm {\d R_{r}} 
  \hskip +0.020 truecm {R^{2}_{s}} \hskip +0.040 truecm {\d R_{s}} \hskip +0.075 truecm
  {\sin \hskip -0.045 truecm {\Theta}_{rs}} \hskip +0.040 truecm {\d {\Theta}_{rs}} \hskip +0.040 truecm {\d {\Phi}_{rs}} 
  \hskip +0.075 truecm {\sin \hskip -0.035 truecm {\theta}_{rs}} \hskip +0.040 truecm 
  {\d {\theta}_{rs}} \hskip +0.040 truecm {\d {\phi}_{rs}},
\label{a3}
\eeq
where $R_{r(s)}$ is the magnitude of the vector ${\vec R}_{r(s)}$. 

The expression of the mean of ${({\e^{ \, i k R_{rs}}}/{k R_{rs}}) \, \e^{ \, i{\vec k}.{\vec R}_{rs}}}$
may be obtained by successive
averages over pairs of variables; the variables with respect to which the integration has  
been done will be indicated in a subscript. The average over ${\Theta}_{rs}$ and ${\Phi}_{rs}$ gives
\begin{eqnarray}
  &&{\left \langle \Biggl (\frac{\e^{ \, i k R_{rs}}}{k R_{rs}} \Biggr ) \,
  \e^{ \, i{\vec k}.{\vec R}_{rs}} \right \rangle}_{({\Theta}_{rs},{\Phi}_{rs})}     
  \nonumber\\
  && = \frac{1}{4 \pi} \, \frac{\e^{ \, i k R_{rs}}}{k R_{rs}}   
  \int_{0}^{\pi} {\sin \hskip -0.045 truecm {\Theta}_{rs}} \hskip +0.040 truecm {\d {\Theta}_{rs}} 
  \int_{0}^{2 \pi} {\d {\Phi}_{rs}} \,
  \e^{ \, i k R_{rs} \hskip -0.016 truecm \cos \hskip -0.018 truecm {\Theta}_{rs}}
  \nonumber\\
  && = \sum_{n=0}^{\infty} \frac{(2i)^{n}}{(n+1)!}(k R_{rs})^{n-1}.
\label{a4}
\end{eqnarray}
It follows that 
\begin{eqnarray}  
  {\left \langle \biggl (\frac{\e^{ \, i k R_{rs}}}{k R_{rs}} \biggr ) \,
  \e^{ \, i{\vec k}.{\vec R}_{rs}} \right \rangle} 
  = \sum_{n=0}^{\infty} \frac{(2i)^{n}}{(n+1)!} \,
  {\left \langle (k R_{rs})^{n-1} \right \rangle}_{({\theta}_{rs},{\phi}_{rs} ; R_{r}, R_{s})}, 
\label{a5} 
\end{eqnarray}
and so we need the expression of the mean of $(k R_{rs})^{n-1}$ for any $n$ to obtain that of the
mean of ${({\e^{ \, i k R_{rs}}}/{k R_{rs}}) \, \e^{ \, i{\vec k}.{\vec R}_{rs}}}$.

The average of $(k R_{rs})^{n-1}$ over the angles gives
\begin{eqnarray}
  && {\left \langle (k R_{rs})^{n-1} \right \rangle}_{({\theta}_{rs},{\phi}_{rs})}
  \nonumber\\
  && = \frac{k^{n-1}}{4 \pi} \int_{0}^{\pi} 
  {\sin \hskip -0.035 truecm {\theta}_{rs}} \hskip +0.040 truecm {\d {\theta}_{rs}} 
  \int_{0}^{2 \pi} {\d {\phi}_{rs}} \,
  (R^{2}_{r} + R^{2}_{s} - 2 R_{r} R_{s} \cos \hskip -0.0129 truecm {\theta}_{rs})^{ \hskip -0.027 truecm \frac{n-1}{2}}
  \nonumber\\
  \nonumber\\
  && = \frac{k^{n-1}}{2 (n+1)} \hskip +0.025 truecm \Biggl ( \frac{(R_{r} + R_{s})^{n+1} - |R_{r} - R_{s}|^{n+1}}
  {R_{r} R_{s}} \Biggr ) \, \, \, \, \, \, \, \, \, \, \, \, \, \, \, \, \, (n = 0,1,...). 
\label{a6}
\end{eqnarray}
The average of the obtained expression gives that of the mean of $(k R_{rs})^{n-1}$. We find  
\begin{eqnarray}
  && {\left \langle (k R_{rs})^{n-1} \right \rangle}_{({\theta}_{rs},{\phi}_{rs} ; R_{r}, R_{s})} 
  \nonumber\\
  && = {\left \langle 
  \frac{k^{n-1}}{2 (n+1)} \hskip +0.025 truecm \Biggl ( \frac{(R_{r} + R_{s})^{n+1} - |R_{r} - R_{s}|^{n+1}}
  {R_{r} R_{s}} \Biggr ) 
  \right \rangle}_{(R_{r}, R_{s})}
  \nonumber\\  
  && = \frac{9} {{\rm R}^{6}} \, \frac{k^{n-1}}{2 (n+1)}
  \int_{0}^{\rm R} {R_{r}} \hskip +0.040 truecm {\d R_{r}} 
  \int_{0}^{\rm R} {R_{s}} \hskip +0.040 truecm {\d R_{s}} \, \lbrack (R_{r} + R_{s})^{n+1} 
  - |R_{r} - R_{s}|^{n+1} \hskip -0.010 truecm \rbrack 
  \nonumber\\ 
  \nonumber\\ 
  && = \frac{72 \, (2 k {\rm R})^{n-1}}{(n+2)(n+3)(n+5)} \, \, \, \, \, \, \, \, \, \, \, \, \, \, \, \, \, \,                                                          
  \, \, \, \, \, \, \, \, \, \, \, \, \, \, \, \, \, \, \, \, \, \, \, \,
  \, \, \, \, \, \, \, \, \, \, \, \, \, \, \, \, \, \, \, \, \, \, \, \, (n = 0,1,...).
\label{a7} 
\end{eqnarray}
Substituting the expression of the mean of $(k R_{rs})^{n-1}$ into~(\ref{a5}), we obtain
\begin{eqnarray} 
  && {\left \langle \Biggl (\frac{\e^{ \, i k R_{rs}}}{k R_{rs}} \Biggr ) \,
  \e^{ \, i{\vec k}.{\vec R}_{rs}} \right \rangle}    
  \nonumber\\
  && = \frac{72}{2 k {\rm R}} \sum_{n=0}^{\infty} \frac{(4 i k {\rm R})^{n}}{(n+1)!(n+2)(n+3)(n+5)}
  \nonumber\\
  && = \frac{144}{(4 k {\rm R})^{5}} \biggl \lbrack \hskip +0.024 truecm \e^{ \, 4 i k {\rm R}} - 1 - 4ik{\rm R} 
  - \frac{(4 i k {\rm R})^{2}}{2} - \frac{(4 i k {\rm R})^{3}}{6} \biggr \rbrack
  \nonumber\\  
  && \, \, \, + i \frac{144}{(4 k {\rm R})^{6}} \biggl \lbrack \hskip +0.024 truecm \e^{ \, 4 i k {\rm R}} - 1 - 4ik{\rm R} 
  - \frac{(4 i k {\rm R})^{2}}{2} - \frac{(4 i k {\rm R})^{3}}{6} - \frac{(4 i k {\rm R})^{4}}{24} \biggr \rbrack .
\label{a8} 
\end{eqnarray}

Substitution of the expression of the mean of  
$(\e^{ \, i k R_{rs}}/{k R_{rs}}) \, \e^{ \, i{\vec k}.{\vec R}_{rs}}$ into~(\ref{a1}) leads 
to that of the mean cross section in the approximation of single and double scattering, 
which is
\begin{eqnarray}    
  {\left \langle {\sigma}_{{\rm s} + {\rm d}} \right \rangle} = {\rm N} {\sigma}    
  + \frac{9 {\rm N} ({\rm N} -1)}{4(k {\rm R})^{6}} \biggl \lbrack \hskip +0.024 truecm
  \sin (2 {\delta}) \hskip +0.060 truecm
  {\rm C}(k {\rm R}) 
  + \cos \hskip +0.022 truecm (2 {\delta}) \, {\rm S}(k {\rm R}) \hskip -0.011 truecm \biggr \rbrack {\sigma},  
\label{a9}
\end{eqnarray}
with
\beq
  {\rm C}(k {\rm R}) = \frac{(k {\rm R}) \cos \hskip +0.022 truecm (4 k {\rm R})}{16} - \frac{\sin (4 k {\rm R})}{64} 
  + \frac{(k {\rm R})^{3}}{3} 
\label{a10}
\eeq
and
\beq 
  {\rm S}(k {\rm R}) = \frac{(k {\rm R}) \sin (4 k {\rm R})}{16} + \frac{\cos \hskip +0.022 truecm (4 k {\rm R})}{64} 
  + \frac{(k {\rm R})^{4}}{2} - \frac{(k {\rm R})^{2}}{8} - \frac{1}{64}.
\label{a11}
\eeq
The functions ${\rm C}(k {\rm R})$ and ${\rm S}(k {\rm R})$ take on only positive values. It is to be noted that the
mean cross section is not bounded in the approximation of single and double scattering
because the function ${\rm C}(k {\rm R})/(k {\rm R})^{6}$ behaves like $1/k {\rm R}$ when ${\rm R}$ tends to $0$.

\appendix
\setcounter{equation}{0}
\def\theequation{B.\arabic{equation}}
\def\thesubsection{B.\arabic{subsection}}
\section*{Appendix B: Expression of the mean cross section for an unlimited number of contributions of successive scatterings}

This appendix is devoted to the derivation of the expression of the mean cross section 
for an unlimited number of contributions of successive scatterings to the cross section.
The relevant expression for the cross section is obtained from~(\ref{e46}) by discarding the 
remainder term and letting the integer ${\rm M}$ become infinitely large. The formula giving the mean
cross section is then  
\begin{eqnarray}  
  {\left \langle {\sigma}_{\rm tot} \right \rangle} && = {\rm N} {\sigma}  
  \nonumber\\   
  && \, \, \, + \frac{4 \pi}{k^{2}} \sum_{{\rm m} = 2}^{\rm M} \, (\sin \hskip -0.035 truecm {\delta})^{\rm m}
  \hskip +0.010 truecm \Im \Biggl \lbrace \e^{ \, i {\rm m} {\delta}}
  \sum_{r_{1} \hskip -0.015 truecm = 1}^{\rm N} 
  \hskip -0.11 truecm
  \sum_{\scriptstyle r_{2} = 1 \atop \scriptstyle (r_{2} \neq r_{1} \hskip -0.015 truecm )}^{\rm N} \dots  
  \sum_{\scriptstyle r_{\rm m} = 1 \atop \scriptstyle (r_{\rm m} \neq r_{{\rm m}-1} \hskip -0.015 truecm )}^{\rm N}
  \hskip -0.11 truecm
  \Biggl \langle \Biggl (\frac{\e^{ \, i k R_{r_{1} \hskip -0.025 truecm r_{2}}}}
  {k R_{r_{1} \hskip -0.025 truecm r_{2}}} \Biggr )
  \nonumber\\
  && \, \, \, \, \, \, \, \,              
  \times \Biggl (\frac{\e^{ \, i k R_{r_{2}r_{3}}}}{k R_{r_{2}r_{3}}} \Biggr )  
  \dots \Biggl (\frac{\e^{ \, i k R_{r_{{\rm m}-1} \hskip -0.025 truecm r_{\rm m}}}}
  {k R_{r_{{\rm m}-1} \hskip -0.025 truecm r_{\rm m}}} \Biggr ) \,  
  \e^{ \, i{\vec k}.{\vec R}_{r_{1} \hskip -0.025 truecm r_{\rm m}}} \hskip -0.070 truecm  \Biggr \rangle \Biggr \rbrace, 
\label{b1}
\end{eqnarray}  
where ${\rm M}$ is as large as desired. Since ${\sigma} = (4 {\pi}/k^{2}) \sin^{2} \hskip -0.045 truecm {\delta}$,   
this formula may be written as
\begin{eqnarray}  
  {\left \langle {\sigma}_{\rm tot} \right \rangle} = \frac{4 \pi}{k^{2}} \hskip +0.035 truecm {\rm N}
  \sum_{{\rm m} = 1}^{\rm M} \, ({\rm N}-1)^{{\rm m}-1} (\sin \hskip -0.035 truecm {\delta})^{\rm m}
  \hskip +0.010 truecm \Im \biggl \lbrace \e^{ \, i {\rm m} {\delta}} M_{\rm m}(k {\rm R}) \biggr \rbrace,    
\label{b2}
\end{eqnarray} 
with
\beq 
  M_{1}(k {\rm R}) = 1
\label{b3}
\eeq
and
\begin{eqnarray}  
  M_{\rm m}(k {\rm R}) && = \frac{1}{{\rm N} ({\rm N} -1)^{{\rm m}-1}} \Biggl \lbrack \,
  \sum_{r_{1} \hskip -0.015 truecm = 1}^{\rm N} 
  \hskip -0.11 truecm
  \sum_{\scriptstyle r_{2} = 1 \atop \scriptstyle (r_{2} \neq r_{1} \hskip -0.015 truecm )}^{\rm N} \dots  
  \sum_{\scriptstyle r_{\rm m} = 1 \atop \scriptstyle (r_{\rm m} \neq r_{{\rm m}-1} \hskip -0.015 truecm )}^{\rm N}
  \hskip -0.11 truecm
  \Biggl \langle \Biggl (\frac{\e^{ \, i k R_{r_{1} \hskip -0.025 truecm r_{2}}}}
  {k R_{r_{1} \hskip -0.025 truecm r_{2}}} \Biggr )
  \nonumber\\
  && \, \, \, \, \, \, \, \, \, \, \, \, \, \, \, \, \, \, \, \, \, \, \, \, \, \, \, \, \, \, \, \, \, \, \,
  \times \Biggl (\frac{\e^{ \, i k R_{r_{2}r_{3}}}}{k R_{r_{2}r_{3}}} \Biggr )  
  \dots \Biggl (\frac{\e^{ \, i k R_{r_{{\rm m}-1} \hskip -0.025 truecm r_{\rm m}}}}
  {k R_{r_{{\rm m}-1} \hskip -0.025 truecm r_{\rm m}}} \Biggr ) \,  
  \e^{ \, i{\vec k}.{\vec R}_{r_{1} \hskip -0.025 truecm r_{\rm m}}} \hskip -0.070 truecm \Biggr \rangle \Biggr \rbrack   
  \nonumber\\
  \nonumber\\
  && \, \, \, \, \, \, \, \, \, \, \, \, \, \, \, \, \, \, \, \, \, \, \, \, \, \, \, \, \, \, \, \, \, \, \, \, 
  \, \, \, \, \, \, \, \, \, \, \, \, \, \, \, \, \, \, \, \, \, \, \, \, \, \, \, \, \, \, \, \, \, \, \, \, \,   
  \, \, \, \, \, \, \, \, \, \, \, \, \, \, \, \, \, \, \, \, \, \, \, \, \, \, \, \, \, \, \, \, \, \,
  ({\rm m} = 2,3 \hskip +0.015 truecm ,...).
\label{b4}
\end{eqnarray} 
Since each scatterer has an equal probability of being at any position inside a sphere of volume ${\rm V}$ 
and radius ${\rm R}$ (whose center is taken as the origin of coordinates), we have
\begin{eqnarray}    
  M_{\rm m}(k {\rm R}) && = \frac{1}{{\rm V}^{\rm m}} 
  \int_{{\rm V}} {\d {\vec R}_{1}} 
  \int_{{\rm V}} {\d {\vec R}_{2}}
  \dots 
  \int_{{\rm V}} {\d {\vec R}_{\rm m}} \, 
  \Biggl (\frac{\e^{ \, i k R_{12}}}{k R_{12}} \Biggr )
  \nonumber\\
  && \, \, \, \, \, \, \, \, \times \Biggl (\frac{\e^{ \, i k R_{23}}}{k R_{23}} \Biggr )  
  \dots \Biggl (\frac{\e^{ \, i k R_{{{\rm m}-1}{\rm m}}}}{k R_{{{\rm m}-1}{\rm m}}} \Biggr ) \,   
  \e^{ \, - i{\vec k}.{\vec R}_{1}} \,
  \e^{ \, i{\vec k}.{\vec R}_{\rm m}}          
  \nonumber\\
  \nonumber\\
  && \, \, \, \, \, \, \, \, \, \, \, \, \, \, \, \, \, \, \, \, \, \, \, \, \, \, \, \, \, \, \, \, \, \, \, \, 
  \, \, \, \, \, \, \, \, \, \, \, \, \, \, \, \, \, \, \, \, \, \, \, \, \, \, \, \, \, \, \, \, \, \, \, \, \,   
  \, \, \, \, \, \, \, \, \, \, \, \, \, \, \, \, \, \, \, \, \, \, \, \, \, \, \, \, \, \, \, \, \, \, 
  ({\rm m} = 2,3,...).
\label{b5}
\end{eqnarray}
Introducing the reduced vectors 
\beq
  {\vec u}_{i} = {{\vec R}_{i}}/ \hskip +0.005 truecm {\rm R} 
  \, \, \, \, \, \, \, \, \, \, \, \, \, \, \, \, \, \, \, \, \, \, \, \, \, \, \, \, \, \, \, \, \, \, \, 
  \, \, \, \, \, \, \, \, \, \, \, \, \, \, \, \, \, \, \, \, \, \, \, \, \, \, \, \, \, \, \, \, \, \, \, 
  \, \, \, \, \, \, \, \, \, \, \, \, \, \, \, \, \, \, \, \, \, \, \, \, \, \, \, \, \, \, \, \, \, \, \,  
  (i = 1 \hskip -0.015 truecm ,2 \hskip +0.015 truecm ,...),
\label{b555}
\eeq
we may also write the expression of $M_{\rm m}(k {\rm R})$ as 
\begin{eqnarray}  
  M_{\rm m}(k {\rm R}) && = \frac{3^{\rm m}}{(4 \pi)^{\rm m}}
  \int {\d {\vec u}_{1}} 
  \int {\d {\vec u}_{2}}
  \dots 
  \int {\d {\vec u}_{\rm m}} \, 
  \Biggl (\frac{\e^{ \, i k {\rm R} |{\vec u}_{2}-{\vec u}_{1}|}}{k {\rm R} |{\vec u}_{2}-{\vec u}_{1}|} \Biggr )
  \nonumber\\
  && \, \, \, \, \, \, \, \, \times \Biggl (\frac{\e^{ \, i k {\rm R} |{\vec u}_{3}-{\vec u}_{2}|}}
  {k {\rm R} |{\vec u}_{3}-{\vec u}_{2}|} \Biggr )  
  \dots \Biggl (\frac{\e^{ \, i k {\rm R} |{\vec u}_{\rm m}-{\vec u}_{{\rm m}-1}|}}
  {k {\rm R} |{\vec u}_{\rm m}-{\vec u}_{{\rm m}-1}|} \Biggr ) \, 
  \e^{ \, - i {\rm R} {\vec k}.{\vec u}_{1}} \,
  \e^{ \, i {\rm R}{\vec k}.{\vec u}_{\rm m}}          
  \nonumber\\
  \nonumber\\
  && \, \, \, \, \, \, \, \, \, \, \, \, \, \, \, \, \, \, \, \, \, \, \, \, \, \, \, \, \, \, \, \, \, \, \, \, 
  \, \, \, \, \, \, \, \, \, \, \, \, \, \, \, \, \, \, \, \, \, \, \, \, \, \, \, \, \, \, \, \, \, \, \, \, \,   
  \, \, \, \, \, \, \, \, \, \, \, \, \, \, \, \, \, \, \, \, \, \, \, \, \, \, \, \, \, \, \, \, \, \, 
  ({\rm m} = 2,3,...),
\label{b444}
\end{eqnarray}
where each integration is over all points inside a sphere of radius 1. The mean of the contribution of any number of successive 
scatterings may be written in the form
\beq
  M_{{\rm n}+1}(k {\rm R}) = \int_{0}^{1} {\d u} \, {\cal P}_{\rm n}(u,k {\rm R})                  
  \, \, \, \, \, \, \, \, \, \, \, \, \, \, \, \, \, \, \, \, \, \, \, \, \, \, \, \, \, \, \, \, \, \, \, \, \, \,
  \, \, \, \, \, \, \, \, \, \, \, \, \, \, \, \, \, \, ({\rm n} = 0,1,...)
\label{b655}
\eeq 
with
\beq
 {\cal P}_{0}(u, k {\rm R}) = 3 \, u^{2},
\label{b6}
\eeq 
\beq  
  {\cal P}_{1}(u, k {\rm R}) = 3 u^{2} \frac{3}{(4 \pi)^{2}} 
  \int {\d {\Omega}_{\vec u}} \, \e^{ \, i {\rm R} {\vec k}.{\vec u}}
  \int {\d {\vec r}_{1}} \, \e^{ \, - i {\rm R} {\vec k}.{\vec u}_{1}}
  \Biggl (\frac{\e^{ \, i k {\rm R} |{\vec u} - {\vec u}_{1} \hskip -0.015 truecm |}}
  {k {\rm R} |{\vec u} - {\vec u}_{1} \hskip -0.025 truecm |} \Biggr ),    
\label{b7}
\eeq  
and
\begin{eqnarray}  
  {\cal P}_{\rm n}(u,k {\rm R}) && = 3 \, u^{2} \frac{3^{\rm n}}{(4 \pi)^{{\rm n}+1}}  
  \int {\d {\Omega}_{\vec u}}
  \int {\d {\vec u}_{1}} \int {\d {\vec u}_{2}}
  \dots \int {\d {\vec u}_{\rm n}} \, 
  \Biggl (\frac{\e^{ \, i k {\rm R}|{\vec u}_{2}-{\vec u}_{1}|}}{k {\rm R}|{\vec u}_{2}-{\vec u}_{1}|} \Biggr )
  \nonumber\\
  && \, \, \, \, \, \, \, \, \times \Biggl (\frac{\e^{ \, i k {\rm R}|{\vec u}_{3}-{\vec u}_{2}|}} 
  {k {\rm R}|{\vec u}_{3}-{\vec u}_{2}|} \Biggr ) 
  \dots \Biggl (\frac{\e^{ \, i k {\rm R}|{\vec u}_{\rm n}-{\vec u}_{{\rm n}-1}|}}
  {k {\rm R}|{\vec u}_{\rm n}-{\vec u}_{{\rm n}-1}|} \Biggr ) \,   
  \e^{ \, - i {\rm R}{\vec k}.{\vec u}_{1}} 
  \nonumber\\
  && \, \, \, \, \, \, \, \, \times \Biggl (\frac{\e^{ \, i k {\rm R}|{\vec u}-{\vec u}_{\rm n}|}}
  {k {\rm R} |{\vec u}-{\vec u}_{\rm n}|} \Biggr ) \,
  \e^{ \, i {\rm R} {\vec k}.{\vec u}}  
  \nonumber\\
  \nonumber\\
  && \, \, \, \, \, \, \, \, \, \, \, \, \, \, \, \, \, \, \, \, \, \, \, \, \, \, \, \, \, \, \, \, \, \, \, \, 
  \, \, \, \, \, \, \, \, \, \, \, \, \, \, \, \, \, \, \, \, \, \, \, \, \, \, \, \, \, \, \, \, \, \, \, \, \,
  \, \, \, \, \, \, \, \, \, \, \, \, \, \, \, \, \, \, \, \, \, \, \, \, \, \, \, \, \, \,  
  ({\rm n} = 2,3,...).
\label{b8}
\end{eqnarray}  
The recursion formula for the quantities ${\cal P}_{\rm n}(u,k {\rm R})$ is then obtained by substituting 
the expansions~\cite{j}
\beq
  \e^{ \, i{\vec k}.{\vec R}} = {4 \pi} \sum_{l=0}^{\infty} \sum_{m=-l}^{l} i^{l} \,
  j_{l}(k R) \, Y_{l m}^{*}({\Omega}_{{\vec k}}) \, Y_{l m}({\Omega}_{{\vec R}}),
\label{b9}
\eeq  
where $R$ and ${\Omega}_{{\vec R}}$ are the magnitude and direction of the vector ${\vec R}$,  
and 
\begin{eqnarray}
  && \frac{\e^{ \, i k {\rm R} |{\vec u}_{s}-{\vec u}_{r}|}}{k {\rm R} |{\vec u}_{s}-{\vec u}_{r}|}                
  \nonumber\\
  && = {4 \pi} i \sum_{l=0}^{\infty} \sum_{m=-l}^{l}
  j_{l}(k {\rm R}{\rm Min}(u_{r},u_{s})) \, h_{l}^{(1)}(k {\rm R}{\rm Max}(u_{r},u_{s})) \, Y_{l m}^{*}({\Omega}_{{\vec u}_{r}}) \, 
  Y_{l m}({\Omega}_{{\vec u}_{s}})
  \nonumber\\
  && \, \, \, \, \, \, \, \, \, \, \, \, \, \, \, \, \, \, \, \, \, \, \, \, \, \, \, \, \, \, \, \, \, \, \, \, \, 
  \, \, \, \, \, \, \, \, \, \, \, \, \, \, \, \, \, \, \, \, \, \, \, \, \, \, \, \, \, \, \, \, \, \, \, \, \, \, 
  \, \, \, \, \, \, \, \, \, \, \, \, \, \, \, \, \, \, \, \, \, \, \, \, \, \, \, \, \, \, \, \, \, \,   
  \, \, \, \, \, \, \, \, \, \, \, \, \, \, \, \, \, \, \, \, \, \, \, \, \, \, \, \, \, \, \, \, (s \neq r),
\label{b10}
\end{eqnarray}  
where $u_{r(s)}$ and ${\Omega}_{{\vec u}_{r(s)}}$ are the magnitude and direction of the vector ${\vec u}_{r(s)}$ 
and where ${\rm Min}(u_{r},u_{s})$ is the smaller and ${\rm Max}(u_{r},u_{s})$ the larger of $u_{r}$ and $u_{s}$,
into~(\ref{b8}). Using the notations ${\Omega}_{{\vec u}_{r}} = {\Omega}_{r}$ $(r=1, ..., {\rm n})$, we obtain 
\begin{eqnarray}
  {\cal P}_{\rm n}(u,k {\rm R}) && = 3 \, u^{2} \, (4 \pi) \, i^{\rm n}
  \sum_{l=0}^{\infty} \sum_{m=-l}^{l} i^{l} \, j_{l}(k {\rm R}) \, Y_{l m}^{*}({\Omega}_{{\vec k}})
  \nonumber\\
  && \, \, \, \, \, \, \, \, \sum_{l'=0}^{\infty} \sum_{m'=-l'}^{l'} (-i)^{l'} \, Y_{l' m'}({\Omega}_{{\vec k}})  
  \sum_{l_{1}=0}^{\infty} \sum_{m_{1}=-l_{1}}^{l_{1}} 
  \sum_{l_{2}=0}^{\infty} \sum_{m_{2}=-l_{2}}^{l_{2}} 
  \dots
  \sum_{l_{\rm n}=0}^{\infty} \sum_{m_{\rm n}=-l_{\rm n}}^{l_{\rm n}} 
  \nonumber\\
  && \, \, \, \, \, \, \, \, \times \int_{0}^{1} \hskip -0.050 truecm 3 \, u_{1}^{2} \hskip +0.010 truecm {\d u_{1}}
  \int_{0}^{1} \hskip -0.050 truecm 3 \, u_{2}^{2} \hskip +0.015 truecm {\d u_{2}}
  \dots 
  \int_{0}^{1} \hskip -0.050 truecm 3 \, u_{\rm n}^{2} \hskip +0.020 truecm {\d u_{\rm n}} \,
  j_{l_{1}}(k {\rm R}{\rm Min}(u_{1},u_{2})) \, 
  \nonumber\\  
  && \, \, \, \, \, \, \, \, \times  h_{l_{1}}^{(1)}(k {\rm R}{\rm Max}(u_{1},u_{2})) \, j_{l_{2}}(k {\rm R}{\rm Min}(u_{2},u_{3})) \, 
  \nonumber\\ 
  && \, \, \, \, \, \, \, \, \times h_{l_{2}}^{(1)}(k {\rm R}{\rm Max}(u_{2},u_{3})) 
  \dots j_{l_{{\rm n}-1}}(k {\rm R}{\rm Min}(u_{{\rm n}-1},u_{\rm n})) 
  \nonumber\\ 
  && \, \, \, \, \, \, \, \, \times 
  h_{l_{{\rm n}-1}}^{(1)}(k {\rm R}{\rm Max}(u_{{\rm n}-1},u_{\rm n})) \, j_{l_{\rm n}}(k {\rm R}{\rm Min}(u_{\rm n},k r))
  \nonumber\\ 
  && \, \, \, \, \, \, \, \, \times 
  h_{l_{\rm n}}^{(1)}(k {\rm R}{\rm Max}(u_{\rm n},k r)) \, j_{l'}(k {\rm R} u_{1}) \int \hskip -0.020 truecm {\d {\Omega}_{\vec r}} 
  \int \hskip -0.020 truecm {\d {\Omega}_{1}}
  \nonumber\\
  && \, \, \, \, \, \, \, \, \times 
  \int \hskip -0.020 truecm {\d {\Omega}_{2}}
  \dots
  \int \hskip -0.020 truecm {\d {\Omega}_{\rm n}} \, 
  Y_{l_{1} m_{1}}^{*}({\Omega}_{1}) \,
  Y_{l_{1} m_{1}}({\Omega}_{2}) \,  
  \nonumber\\
  && \, \, \, \, \, \, \, \, \times  Y_{l_{2} m_{2}}^{*}({\Omega}_{2}) \, Y_{l_{2} m_{2}}({\Omega}_{3}) \dots
  Y_{l_{{\rm n}-1} m_{{\rm n}-1}}^{*}({\Omega}_{{\rm n}-1}) \, Y_{l_{{\rm n}-1} m_{{\rm n}-1}}({\Omega}_{\rm n})
  \nonumber\\
  && \, \, \, \, \, \, \, \, \times 
  Y_{l_{\rm n} m_{\rm n}}^{*}({\Omega}_{\rm n}) \,
  Y_{l_{\rm n} m_{\rm n}}({\Omega}_{\vec r}) \, Y_{l m}({\Omega}_{\vec r}) \,
  Y_{l' m'}({\Omega}_{1}).
  \nonumber\\ 
\label{b11}
\end{eqnarray}
Use of the orthogonality relation for the spherical harmonics and of the identity~\cite{j}
\beq 
  \sum_{m=-l}^{l} Y_{l m}^{*}({\Omega}_{{\vec k}}) \, Y_{l m}({\Omega}_{{\vec k}})
  = \frac{2l+1}{4 \pi}
\label{b12}
\eeq
then leads to 
\begin{eqnarray}
  {\cal P}_{\rm n}(u,k {\rm R}) && = 3 \, u^{2} \, i^{\rm n}  
  \sum_{l=0}^{\infty} (2l+1) \, j_{l}(k {\rm R} u)  
  \nonumber\\ 
  && \, \, \, \, \, \, \, \, \times \int_{0}^{1} \hskip -0.050 truecm 3 \, u_{1}^{2} \hskip +0.010 truecm {\d u_{1}}
  \int_{0}^{1} \hskip -0.050 truecm 3 \, u_{2}^{2} \hskip +0.015 truecm {\d u_{2}}
  \dots 
  \int_{0}^{1} \hskip -0.050 truecm 3 \, u_{\rm n}^{2} \hskip +0.020 truecm {\d u_{\rm n}} \,
  j_{l}(k {\rm R}{\rm Min}(u_{1},u_{2})) 
  \nonumber\\  
  && \, \, \, \, \, \, \, \, \times h_{l}^{(1)}(k {\rm R}{\rm Max}(u_{1},u_{2}))  
  j_{l}(k {\rm R}{\rm Min}(u_{2},u_{3})) \, 
  \nonumber\\ 
  && \, \, \, \, \, \, \, \, \times
  h_{l}^{(1)}(k {\rm R}{\rm Max}(u_{2},u_{3}))  
  \dots
  j_{l}(k {\rm R}{\rm Min}(u_{{\rm n}-1},u_{\rm n}))  
  \nonumber\\ 
  && \, \, \, \, \, \, \, \, \times  
  h_{l}^{(1)}(k {\rm R}{\rm Max}(u_{{\rm n}-1},u_{\rm n})) \,
  j_{l}(k {\rm R}{\rm Min}(u_{\rm n},u))
  \nonumber\\ 
  && \, \, \, \, \, \, \, \, \times  
  h_{l}^{(1)}(k {\rm R}{\rm Max}(u_{\rm n},u)) \, j_{l}(k {\rm R} u_{1})  
  \nonumber\\ 
  && \, \, \, \, \, \, \, \, \, \, \, \, \, \, \, \, \, \, \, \, \, \, \, \, \, \, \, \, \, \, \, \, \, \, \,
  \, \, \, \, \, \, \, \, \, \, \, \, \, \, \, \, \, \, \, \, \, \, \, \, \, \, \, \, \, \, \, \, \, \, \, \,
  \, \, \, \, \, \, \, \, \, \, \, \, \, \, \, \, \, \, \, \, \, \, \, \, \, \, \, \, \, \, \, \, 
  ({\rm n} = 2,3,...)
\label{b13}
\end{eqnarray}
and 
\begin{eqnarray}
  {\cal P}_{1}(u,k {\rm R}) && = 3 \, u^{2} \, i 
  \sum_{l=0}^{\infty} (2l+1) \, j_{l}(k {\rm R} u)  
  \nonumber\\ 
  && \, \, \, \, \, \, \, \, \times \int_{0}^{1} \hskip -0.050 truecm 3 \, u_{1}^{2} \hskip +0.010 truecm {\d u_{1}} \,
  j_{l}(k{\rm R} u_{1}) \, j_{l}(k {\rm R} {\rm Min}(u_{1},u)) \, h_{l}^{(1)}(k {\rm R} {\rm Max}(u_{1},u))  
  \nonumber\\ 
\label{b14}
\end{eqnarray}
We then decompose ${\cal P}_{\rm n}(u,k {\rm R})$ as
\beq 
  {\cal P}_{\rm n}(u,k {\rm R}) = u \sum_{l=0}^{\infty} (2l+1) \, j_{l}(k {\rm R} u) \, 
  {\cal P}_{\rm n}^{(l)}(u, k {\rm R}) \, \, \, \, \, \, \, \, \, \, \, \, \, \, \, \, ({\rm n} = 1,2,...).
\label{b15}
\eeq
We obtain
\begin{eqnarray}
  {\cal P}_{\rm n}^{(l)}(u,k {\rm R}) && = 3 \, u \, i^{\rm n}   
  \int_{0}^{1} \hskip -0.050 truecm 3 \, u_{1}^{2} \hskip +0.010 truecm {\d u_{1}}
  \int_{0}^{1} \hskip -0.050 truecm 3 \, u_{2}^{2} \hskip +0.015 truecm {\d u_{2}}
  \dots 
  \int_{0}^{1} \hskip -0.050 truecm 3 \, u_{\rm n}^{2} \hskip +0.020 truecm {\d u_{\rm n}} \, j_{l}(k {\rm R}u_{1})
  \nonumber\\
  && \, \, \, \, \, \, \, \, \times j_{l}(k {\rm R}{\rm Min}(u_{1},u_{2})) \, h_{l}^{(1)}(k {\rm R}{\rm Max}(u_{1},u_{2})) \,
  j_{l}(k {\rm R}{\rm Min}(u_{2},u_{3}))
  \nonumber\\  
  && \, \, \, \, \, \, \, \, \times h_{l}^{(1)}(k {\rm R}{\rm Max}(u_{2},u_{3}))  
  \dots 
  j_{l}(k {\rm R}{\rm Min}(u_{{\rm n}-1},u_{\rm n})) 
  \nonumber\\
  && \, \, \, \, \, \, \, \, \times h_{l}^{(1)}(k {\rm R}{\rm Max}(u_{{\rm n}-1},u_{\rm n})) \,
  j_{l}(k {\rm R}{\rm Min}(u_{\rm n},u))
  \nonumber\\ 
  && \, \, \, \, \, \, \, \, \times  
  h_{l}^{(1)}(k {\rm R}{\rm Max}(u_{\rm n},u))   
  \nonumber\\ 
  && = 3 \, u \, i^{\rm n} \, h_{l}^{(1)}(k {\rm R} u) 
  \int_{0}^{u} \hskip -0.050 truecm 3 \, u_{\rm n}^{2} \hskip +0.020 truecm {\d u_{\rm n}} \, j_{l}(u_{\rm n})
  \int_{0}^{1} \hskip -0.050 truecm 3 \, u_{1}^{2} \hskip +0.010 truecm {\d u_{1}}
  \int_{0}^{1} \hskip -0.050 truecm 3 \, u_{2}^{2} \hskip +0.015 truecm {\d u_{2}}
  \nonumber\\
  && \, \, \, \, \, \, \, \, \times
  \dots 
  \int_{0}^{1} \hskip -0.050 truecm 3 \, u_{{\rm n}-1}^{2} \hskip +0.020 truecm {\d u_{{\rm n}-1}} \, j_{l}(k {\rm R}u_{1})  
  j_{l}(k {\rm R}{\rm Min}(u_{1},u_{2})) 
  \nonumber\\
  && \, \, \, \, \, \, \, \, \times h_{l}^{(1)}(k {\rm R}{\rm Max}(u_{1},u_{2})) \, j_{l}(k {\rm R}{\rm Min}(u_{2},u_{3})) 
  \nonumber\\  
  && \, \, \, \, \, \, \, \, \times h_{l}^{(1)}(k {\rm R}{\rm Max}(u_{2},u_{3}))  
  \dots j_{l}(k {\rm R}{\rm Min}(u_{{\rm n}-2},u_{{\rm n}-1}))   
  \nonumber\\ 
  && \, \, \, \, \, \, \, \, \times h_{l}^{(1)}(k {\rm R}{\rm Max}(u_{{\rm n}-2},u_{{\rm n}-1}))  
  j_{l}(k {\rm R}{\rm Min}(u_{{\rm n}-1},u_{\rm n})) \, 
  \nonumber\\ 
  && \, \, \, \, \, \, \, \, \times h_{l}^{(1)}(k {\rm R}{\rm Max}(u_{{\rm n}-1},u_{\rm n}))
  \nonumber\\ 
  && \, \, \, + 3 \, u \, i^{\rm n} \, j_{l}(k {\rm R} u)
  \int_{u}^{1} \hskip -0.050 truecm 3 \, u_{\rm n}^{2} \hskip +0.020 truecm {\d u_{\rm n}} \, h_{l}^{(1)}(u_{\rm n})
  \int_{0}^{1} \hskip -0.050 truecm 3 \, u_{1}^{2} \hskip +0.010 truecm {\d u_{1}}
  \int_{0}^{1} \hskip -0.050 truecm 3 \, u_{2}^{2} \hskip +0.015 truecm {\d u_{2}}
  \nonumber\\
  && \, \, \, \, \, \, \, \, \times
  \dots 
  \int_{0}^{1} \hskip -0.050 truecm 3 \, u_{{\rm n}-1}^{2} \hskip +0.020 truecm {\d u_{{\rm n}-1}} \, 
  j_{l}(k {\rm R} \hskip +0.010 truecm u_{1}) \,
  j_{l}(k {\rm R} \hskip +0.015 truecm {\rm Min}(u_{1},u_{2}))  
  \nonumber\\
  && \, \, \, \, \, \, \, \, \times  
  h_{l}^{(1)}(k {\rm R} \hskip +0.020 truecm {\rm Max}(u_{1},u_{2})) \, j_{l}(k {\rm R} \hskip +0.025 truecm {\rm Min}(u_{2},u_{3}))
  \nonumber\\  
  && \, \, \, \, \, \, \, \, \times  
  h_{l}^{(1)}(k {\rm R} \hskip +0.030 truecm {\rm Max}(u_{2},u_{3}))  
  \dots
  j_{l}(k {\rm R} {\rm Min}(u_{{\rm n}-2},u_{{\rm n}-1}))
  \nonumber\\ 
  && \, \, \, \, \, \, \, \, \times  
  h_{l}^{(1)}(k {\rm R} \, {\rm Max}(u_{{\rm n}-2},u_{{\rm n}-1})) \, 
  j_{l}(k {\rm R} \, {\rm Min}(u_{{\rm n}-1},u_{\rm n})) 
  \nonumber\\ 
  && \, \, \, \, \, \, \, \, \times    
  h_{l}^{(1)}(k {\rm R} \, {\rm Max}(u_{{\rm n}-1},u_{\rm n}))   
  \nonumber\\ 
  && = 3i \, u \, h_{l}^{(1)}(k {\rm R} u) 
  \int_{0}^{u} \hskip -0.050 truecm {\d s} \, s \, j_{l}(k {\rm R} s) \, {\cal P}_{{\rm n}-1}^{(l)}(s,k{\rm R})
  \nonumber\\ 
  && \, \, \, + 3i \, u \, j_{l}(k {\rm R} u)
  \int_{u}^{1} \hskip -0.050 truecm {\d s} \, s \, h_{l}^{(1)}(k {\rm R} s) \, {\cal P}_{{\rm n}-1}^{(l)}(s,k{\rm R})   
  \nonumber\\ 
  && \, \, \, \, \, \, \, \, \, \, \, \, \, \, \, \, \, \, \, \, \, \, \, \, \, \, \, \, \, \, \, \, \,
  \, \, \, \, \, \, \, \, \, \, \, \, \, \, \, \, \, \, \, \, \, \, \, \, \, \, \, \, \, \, \, \, \,
  \, \, \, \, \, \, \, \, \, \, \, \, \, \, \, \, \, \, \, \, \, \, \, \, \, \, \, \, \, \, \, \, \, \, 
  ({\rm n} = 2,3,...)
\label{b16}
\end{eqnarray}
and 
\begin{eqnarray}
  {\cal P}_{1}^{(l)}(u,k {\rm R}) && = 3i \, u  
  \int_{0}^{1} \hskip -0.050 truecm 3 \, u_{1}^{2} \hskip +0.010 truecm {\d u_{1}} \, j_{l}(k{\rm R} ~ u_{1})
  j_{l}(k {\rm R} ~ {\rm Min}(u_{1},u)) 
  \nonumber\\
  && \, \, \, \, \, \, \, \, \times  h_{l}^{(1)}(k {\rm R} \, {\rm max}(u_{1},u))  
  \nonumber\\
  && = 3i \, u \, h_{l}^{(1)}(k {\rm R} u) 
  \int_{0}^{u} \hskip -0.050 truecm 3 \, s^{2} \hskip +0.020 truecm {\d s} \, j_{l}(k {\rm R} s) \, j_{l}(s)
  \nonumber\\ 
  && \, \, \, + 3i \, u \, j_{l}(k {\rm R} u)
  \int_{u}^{1} \hskip -0.050 truecm 3 \, s^{2} \hskip +0.020 truecm {\d s} \, h_{l}^{(1)}(k {\rm R} s) \, j_{l}(s)  
  \nonumber\\ 
  && = 3i \, u \, h_{l}^{(1)}(k {\rm R} u) 
  \int_{0}^{u} \hskip -0.050 truecm {\d s} \, s \, j_{l}(k {\rm R} s) \,
  {\cal P}_{0}^{(l)}(s, k {\rm R})
  \nonumber\\ 
  && \, \, \, + 3i \, u \, j_{l}(k {\rm R} u)
  \int_{u}^{1} \hskip -0.050 truecm {\d s} \, s \, 
  h_{l}^{(1)}(k {\rm R} s) \, {\cal P}_{0}^{(l)}(s, k  {\rm R}) ,   
\label{b17}
\end{eqnarray}
with
\beq 
  {\cal P}_{0}^{(l)}(s, k {\rm R}) = 3 \, s \, j_{l}(k {\rm R} \, s).
\label{b18}
\eeq
This definition of ${\cal P}_{0}^{(l)}(u, k{\rm R})$ is compatible with that of ${\cal P}_{0}(u, k {\rm R})$ as given
in~(\ref{b6}) because of the identity  
\beq 
  \sum_{l=0}^{\infty} (2l+1) \, j_{l}^{2}(u) = 1 .
\label{b19}
\eeq
Each function ${\cal P}_{\rm n}^{(l)}(u, k{\rm R})$ satisfies a differential equation which is obtained by deriving
the recursion relation twice with respect to the variable $u$. Using the facts that~\cite{j}
\beq
  j_{l}(z) \frac{d}{d z} h_{l}^{(1)}(z) - h_{l}^{(1)}(z) \frac{d}{d z} j_{l}(z) = \frac{1}{z^{2}} 
  \, \, \, \, \, \, \, \, \, \, \, \, \, \, \, \, \, \, \, \, \, \, \, \, \, \, \, \, \, \, \, \, \,
  \, \, \, (l = 0,1, ...)
\label{b20}
\eeq
and 
\beq
  \frac{d^{2}}{d z^{2}} f_{l}(z) + \frac{2}{z} \frac{d}{d z} f_{l}(z) 
  + \Biggl (1 - \frac{l(l+1)}{z^{2}} \Biggr ) f_{l}(z) = 0
  \, \, \, \, \, \, \, \, \, (l = 0,1, ...)
\label{b21}
\eeq
for both $j_{l}(z)$ and $h_{l}^{(1)}(z)$, we obtain 
\begin{eqnarray}
  \frac{d^{2}}{d u^{2}} {\cal P}_{\rm n}^{(l)}(u, k{\rm R}) && =  - \frac{3}{k{\rm R}} {\cal P}_{{\rm n}-1}^{(l)}(u, k{\rm R})
  - \Biggl ( (k{\rm R})^{2} - \frac{l(l+1)}{u^{2}} \Biggr ) {\cal P}_{\rm n}^{(l)}(u, k{\rm R}) 
  \nonumber\\
  && \, \, \, \, \, \, \, \, \, \, \, \, \, \, \, \, \, \, \, \, \, \, \, \, \, \, \, \, \, \, \,
  \, \, \, \, \, \, \, \, \, \, \, \, \, \, \, \, \, \, \, \, \, \, \, \, \, \, \, \, \, \, \, \,
  \, \, \, \, \, \, \, \, \, \, \, \, \, \, \, \, \, \, \, \, \, \, \, \, \, \, \, \, (n \neq 0)
  \nonumber\\
  && \, \, \, \, \, \, \, \, \, \, \, \, \, \, \, \, \, \, \, \, \, \, \, \, \, \, \, \, \, \, \,
  \, \, \, \, \, \, \, \, \, \, \, \, \, \, \, \, \, \, \, \, \, \, \, \, \, \, \, \, \, \, \, \,
  \, \, \, \, \, \, \, \, \, \, \, \, \, \, \, \, \, \, \, \, \, \, \, \, \, \, \, \, (l = 0,1, ...)
\label{b22}
\end{eqnarray}
and 
\begin{eqnarray}
  \frac{d^{2}}{d u^{2}} {\cal P}_{0}^{(l)}(u, k{\rm R}) && = 
  - \Biggl ( (k{\rm R})^{2} - \frac{l(l+1)}{u^{2}} \Biggr ) {\cal P}_{0}^{(l)}(u, k{\rm R}) 
  \nonumber\\
  && \, \, \, \, \, \, \, \, \, \, \, \, \, \, \, \, \, \, \, \, \, \, \, \, \, \, \, \, \, \, \,
  \, \, \, \, \, \, \, \, \, \, \, \, \, \, \, \, \, \, \, \, \, \, \, \, \, \, \, \, \, \, \, \,
  \, \, \, \, \, \, \, \, \, \, \, \, \, \, \, \, \, \, \, \, \, \, \, \, \, \, \, \, (l = 0,1, ...).
\label{b23}
\end{eqnarray}
Comparison between these last two equations shows that 
\beq
  {\cal P}_{-1}^{(l)}(u, k{\rm R}) = 0 \, \, \, \, \, \, \, \, \, \, \, \, \, \, \, \, \, \, \, \, \, \, \, \, \, \, \,
  \, \, \, \, \, \, \, \, \, \, \, \, \, \, \, \, \, \, \, \, \, \, \, \, \, \, \, \, \, \, \, \, \, \, \, \, \, \, \, 
  \, \, \, \, \, \, \, \, \, \, \, \, \, \, \, \, \, \, \, \, \, \, \, \, (l = 0,1, ...).
\label{b24}
\eeq
We now introduce the generating function 
\beq
  G^{(l)}(x,u,k{\rm R}) = \sum_{{\rm n} = 0}^{\infty} x^{\rm n} {\cal P}_{\rm n}^{(l)}(u, k{\rm R})
\label{b25}
\eeq
It follows from~(\ref{b22}) and~(\ref{b23}) that the generating function satisfies the equation 
\begin{eqnarray}
  \frac{\partial^{2}}{\partial u^{2}} G^{(l)}(x,u,k{\rm R}) && = 
  - \Biggl ( (k{\rm R})^{2} + \frac{3x}{k {\rm R}} - \frac{l(l+1)}{u^{2}} \Biggr ) G^{(l)}(x,u,k{\rm R})
  \nonumber\\
  && \, \, \, \, \, \, \, \, \, \, \, \, \, \, \, \, \, \, \, \, \, \, \, \, \, \, \, \, \, \, \,
  \, \, \, \, \, \, \, \, \, \, \, \, \, \, \, \, \, \, \, \, \, \, \, \, \, \, \, \, \, \, \, \,
  \, \, \, \, \, \, \, \, \, \, \, \, \, \, \, \, \, \, \, \, \, \, (l = 0,1, ...).
\label{b26}
\end{eqnarray}
The mean cross section can be calculated with the help of the generating function. Substituting~(\ref{b15}) 
into~(\ref{b655}) and then~(\ref{b655}) into~(\ref{b2}), we obtain 
\beq
  {\left \langle {\sigma}_{\rm tot} \right \rangle} = \frac{4 \pi}{k^{2}} {\rm N} 
  \sum_{l=0}^{\infty} (2l+1) \, {\left \langle {\sigma}_{\rm tot} \right \rangle}^{(l)},    
\label{b27}
\eeq
with 
\begin{eqnarray}
  && {\left \langle {\sigma}_{\rm tot} \right \rangle}^{(l)}
  \nonumber\\  
  && = \Im \biggl \lbrace \sin {\delta} \, 
  \e^{ \, i {\delta}} \int_{0}^{1} \hskip -0.050 truecm {\d u} \, u \, j_{l}(k {\rm R} \hskip +0.010truecm u) \, 
  \sum_{{\rm n}=0}^{{\rm M}-1} \biggl ( ( {\rm N}-1 ) \sin {\delta} \e^{ \, i {\delta}} \biggr )^{\rm n} \, 
  {\cal P}_{\rm n}^{(l)}(u,k {\rm R}) \biggr \rbrace.  
  \nonumber\\  
\label{b28}
\end{eqnarray}
Taking ${\rm M}$ to be infinitely large and using~(\ref{b25}), we find  
\begin{eqnarray}
  && {\left \langle {\sigma}_{\rm tot} \right \rangle}^{(l)}
  \nonumber\\  
  && = \Im \biggl \lbrace \sin {\delta} \, 
  \e^{ \, i {\delta}} \int_{0}^{1} \hskip -0.050 truecm {\d u} \, u \, j_{l}(k {\rm R} \hskip +0.010truecm u) \, 
  G^{(l)} \biggl (x = ( {\rm N}-1 ) \sin {\delta} \e^{ \, i {\delta}},u k {\rm R} \biggr ) \biggr \rbrace. 
  \nonumber\\  
\label{b29}
\end{eqnarray}
The regular solution of~(\ref{b26}) is 
\beq
  G^{(l)}(x,u,k{\rm R}) = A_{l}(x) \, u \, j_{l} \Biggl ({\sqrt {(k{\rm R})^{2} + \frac{3x}{k{\rm R}}}} \hskip +0.050 truecm u \Biggr )
  \, \, \, \, \, \, \, \, \, \, \, \, \, \, \, \, \, (l = 0,1, ...).
\label{b30}
\eeq
The function $A_{l}(x)$ is found as follows. It follows from~(\ref{b16}) that
\begin{eqnarray}
  && {\cal P}_{\rm n}^{(l)}(u=1, k{\rm R}) = 3i \, h_{l}^{(1)}(k{\rm R}) 
  \int_{0}^{1} \hskip -0.050 truecm {\d s} \, s \, j_{l}(k {\rm R} s) \, {\cal P}_{{\rm n}-1}^{(l)}(s,k{\rm R}),
  \nonumber\\
  && \, \, \, \, \, \, \, \, \, \, \, \, \, \, \, \, \, \, \, \, \, \, \, \, \, \, \, \, \, \, \, \, \, \, \, \, 
  \, \, \, \, \, \, \, \, \, \, \, \, \, \, \, \, \, \, \, \, \, \, \, \, \, \, \, \, \, \, \, \, \, \, \, \, \,
  \, \, \, \, \, \, \, \, \, \, \, \, \, \, \, \, \, \, \, \, \, \, \, \, \, \, \, \, \, \, \, \, \, \, \, \, \,
  \, \, \, \, \, \, \, \, \, \, \, \, \, \, \, \, \, \, \, (l = 0,1, ...).
\label{b31}
\end{eqnarray}
This implies that 
\begin{eqnarray}
  && \sum_{{\rm n}=1}^{\infty} x^{\rm n} \, {\cal P}_{\rm n}^{(l)}(u=1, k{\rm R}) 
  \nonumber\\  
  && = 3i \, h_{l}^{(1)}(k {\rm R})
  \int_{0}^{1} \hskip -0.050 truecm {\d s} \, s \, j_{l}(k {\rm R} s) \, 
  \sum_{{\rm n}=1}^{\infty} x^{\rm n} \, {\cal P}_{{\rm n}-1}^{(l)}(s,k{\rm R}),
  \nonumber\\
  && \, \, \, \, \, \, \, \, \, \, \, \, \, \, \, \, \, \, \, \, \, \, \, \, \, \, \, \, \, \, \, \, \, \, \, \, 
  \, \, \, \, \, \, \, \, \, \, \, \, \, \, \, \, \, \, \, \, \, \, \, \, \, \, \, \, \, \, \, \, \, \, \, \, \,
  \, \, \, \, \, \, \, \, \, \, \, \, \, \, \, \, \, \, \, \, \, \, \, \, \, \, \, \, \, \, \, \, \, \, \, \, \,
  \, \, \, \, \, \, \, \, \, \, \, \, \, \, \, \, \, \, \, (l = 0,1, ...).
\label{b32}
\end{eqnarray}
and so
\begin{eqnarray}
  && G^{(l)}(x,u=1,k{\rm R}) - {\cal P}_{0}^{(l)}(u=1,k{\rm R}) 
  \nonumber\\
  && = 3i \, x \, h_{l}^{(1)}(k {\rm R})
  \int_{0}^{1} \hskip -0.050 truecm {\d s} \, s \, j_{l}(k {\rm R} s) \, G^{(l)}(x,s,k{\rm R}).
  \nonumber\\
  && \, \, \, \, \, \, \, \, \, \, \, \, \, \, \, \, \, \, \, \, \, \, \, \, \, \, \, \, \, \, \, \, \, \, \, \, 
  \, \, \, \, \, \, \, \, \, \, \, \, \, \, \, \, \, \, \, \, \, \, \, \, \, \, \, \, \, \, \, \, \, \, \, \, \,
  \, \, \, \, \, \, \, \, \, \, \, \, \, \, \, \, \, \, \, \, \, \, \, \, \, \, \, \, \, \, \, \, \, \, \, \, \,   
  \, \, \, \, \, \, \, \, \, \, \, \, \, \, \, \, \, \, \, (l = 0,1, ...).
\label{b33}
\end{eqnarray}
Substituting~(\ref{b30}) in the case $u=1$ as well as in the general case and~(\ref{b18}) in the case $s=1$ into 
this equation, we obtain
\begin{eqnarray}
  && A_{l}(x) \, j_{l} \Biggl ({\sqrt {(k{\rm R})^{2} + \frac{3x}{k{\rm R}}}} \Biggr ) - 3 j_{l}(k {\rm R}) 
  \nonumber\\
  && = 3i \, x \, h_{l}^{(1)}(k {\rm R}) \, A_{l}(x)
  \int_{0}^{1} \hskip -0.050 truecm {\d s} \, s^{2} \, j_{l}(k {\rm R} \hskip +0.050 truecm s) \, 
  j_{l} \Biggl ({ \sqrt {(k{\rm R})^{2} + \frac{3x}{k{\rm R}}}} \hskip +0.030 truecm s \Biggr ).                     
  \nonumber\\
  \nonumber\\
  && \, \, \, \, \, \, \, \, \, \, \, \, \, \, \, \, \, \, \, \, \, \, \, \, \, \, \, \, \, \, \, \, \, \, \, \, 
  \, \, \, \, \, \, \, \, \, \, \, \, \, \, \, \, \, \, \, \, \, \, \, \, \, \, \, \, \, \, \, \, \, \, \, \, \,
  \, \, \, \, \, \, \, \, \, \, \, \, \, \, \, \, \, \, \, \, \, \, \, \, \, \, \, \, \, \, \, \, \, \, \, \, \,  
  \, \, \, \, \, \, \, \, \, \, \, \, \, \, \, \, \, \, \, (l = 0,1, ...).
\label{b34}
\end{eqnarray}
Using the fact that
\begin{eqnarray}
  {\cal I}_{l}(x,k{\rm R}) && = \int_{0}^{1} \hskip -0.050 truecm {\d s} \, s^{2} \, j_{l}(k {\rm R} \hskip +0.050 truecm s) \, 
  j_{l} \Biggl ({\sqrt {(k{\rm R})^{2} + \frac{3x}{k{\rm R}}}} \hskip +0.030 truecm s \Biggr )
  \nonumber\\
  && = \frac{k{\rm R}}{3x} \, \Biggl \lbrack {\sqrt {(k{\rm R})^{2} + \frac{3x}{k{\rm R}}}} \, \, j_{l}(k{\rm R}) \, \,
  j_{l+1} \Biggl ({\sqrt {(k{\rm R})^{2} + \frac{3x}{k{\rm R}}}} \Biggr )
  \nonumber\\
  \nonumber\\
  && \, \, \, - (k{\rm R}) \, \, j_{l} \Biggl ({\sqrt {(k{\rm R})^{2} + \frac{3x}{k{\rm R}}}} \Biggr ) \, \, j_{l+1}(k{\rm R}) \Biggr \rbrack
  \nonumber\\
  \nonumber\\
  && \, \, \, \, \, \, \, \, \, \, \, \, \, \, \, \, \, \, \, \, \, \, \, \, \, \, \, \, \, \, \, \, \, \, \, \, 
  \, \, \, \, \, \, \, \, \, \, \, \, \, \, \, \, \, \, \, \, \, \, \, \, \, \, \, \, \, \, \, \, \, \, \, \, \,
  \, \, \, \, \, \, \, \, \, \, \, \, \, \, \, \, \, \, \, \, \, \, \, \, \, \, \, \, \, \, \, \, \, 
  (l = 0,1, ...),
\label{b35}
\end{eqnarray}
we obtain 
\begin{eqnarray}
  A_{l}(x) = \frac{3 \, j_{l}(k{\rm R})}{j_{l}(k {\cal R}) - 3i \, x \, h_{l}^{(1)}(k {\rm R}) \, {\cal I}_{l}(x,k{\rm R})}
  \, \, \, \, \, \, \, \, \, \, \, \, \, \, \, \, \, \, \, \, \, \, \, \, \, \, (l = 0,1, ...),
\label{b36}
\end{eqnarray}
where we have used the notation $k {\cal R} = {\sqrt {(k{\rm R})^{2} + \frac{3x}{k{\rm R}}}}$. Substituting this equation 
into~(\ref{b30}), we obtain the expression of the generating function, which is 
\beq
  G^{(l)}(x,u,k{\rm R}) = \frac{3u \, j_{l}(k{\rm R}) \, j_{l}(k {\cal R} \hskip +0.010truecm u)}
  {j_{l}(k {\cal R}) - 3i \, x \, h_{l}^{(1)}(k {\rm R}) \, {\cal I}_{l}(x,k{\rm R})} \, \, \, \, \, \, \, (l = 0,1, ...).
\label{b37}
\eeq
Substituting this equation into~(\ref{b29}), we find  
\begin{eqnarray}
  && {\left \langle {\sigma}_{\rm tot} \right \rangle}^{(l)}
  \nonumber\\  
  && = \Im \Biggl \lbrace \frac{3 \sin {\delta} \, \e^{ \, i {\delta}} \, j_{l}(k{\rm R}) \, 
  {\cal I}_{l}(x = ( {\rm N}-1 ) \sin {\delta} \e^{ \, i {\delta}},k{\rm R})}
  {j_{l}(k {\cal R}) 
  - 3i \, ( {\rm N}-1 ) \sin {\delta} \e^{ \, i {\delta}} \, h_{l}^{(1)}(k{\rm R}) \, 
  {\cal I}_{l}(x = ( {\rm N}-1 ) \sin {\delta} \e^{ \, i {\delta}},k{\rm R})} \Biggr \rbrace.  
  \nonumber\\  
  \nonumber\\ 
\label{b38}
\end{eqnarray}
Using~(\ref{b35}) in this equation and then using~(\ref{b27}), we obtain the expression of the mean cross 
section for an unlimited number of contributions of successive scatterings. We find
\begin{eqnarray}
  && \Biggl (\frac{k^{2}}{4 \pi} \Biggr ) \Biggl (\frac{{{\rm N}-1}}{{\rm N}} \Biggr )
  {\left \langle {\sigma}_{\rm tot} \right \rangle} 
  \nonumber\\
  && = \sum_{l=0}^{\infty} (2l+1) \, 
  \Im \Biggl \lbrace i \Biggl \lbrack \frac{({\cal K} {\rm R}) \, j_{l}(k{\rm R}) \, j_{l+1}({\cal K} {\rm R})
  - (k {\rm R}) \, j_{l}({\cal K} {\rm R}) \, j_{l+1}(k{\rm R})}
  {({\cal K} {\rm R}) \,  h_{l}^{(1)}(k{\rm R}) \, j_{l+1}({\cal K} {\rm R}) - (k {\rm R}) \, j_{l}({\cal K} {\rm R}) \, h_{l+1}^{(1)}(k{\rm R})}
  \Biggr \rbrack \Biggr \rbrace,     
  \nonumber\\
  \nonumber\\ 
\label{b39}
\end{eqnarray}
with 
\beq
  {\cal K} = k \, \sqrt {1 + \frac{3 ( {\rm N}-1 ) \sin \hskip -0.035 truecm {\delta} \, \e^{ \, i
  {\delta}}}{(k{\rm R})^{3}}}.
\label{b40}
\eeq


\newpage

\Bibliography{99}

\bibitem{rn}
See, e.g., van Rossum M C W and Nieuwenhuizen Th M 1999 {\it Rev. Mod. Phys.} {\bf 71} 313 and references therein

\bibitem{vcl}
See, e.g., de Vries P, van Coevorden D V and Lagendijk A 1998 {\it Rev. Mod. Phys.} {\bf 70} 447 and references therein

\bibitem{lt}
Lagendijk A and van Tiggelen B A 1996 {\it Phys. Rep.} {\bf 270} 143 and references therein

\bibitem{ts}
van Tiggelen B A and Skipetrov S E (eds.) 2003 {\it Wave Scattering in Complex Media: From Theory to Applications}
(Dordrecht: Kluwer) and references therein

\bibitem{am}
Akkermans E and Montambaux G 2007 {\it Mesoscopic Physics of Electrons and Photons} 
(Cambridge: Cambridge University Press) and references therein

\bibitem{ln}
Luck J M and Nieuwenhuizen Th M 1999 {\it Eur. Phys. J. B} {\bf 7} 483

\bibitem{gw}
See, e.g., Goldberger M L and Watson K M 1964 {\it Collision Theory} (John Wiley~\& Sons)

\bibitem{ajp}
See, e.g., Austern N, Tabakin F and Silver M 1977 {\it Am. J. Phys.} {\bf 45} 361

\bibitem{w}
Watson K M 1953 {\it Phys. Rev.} {\bf 89} 575

\bibitem{s}
See, e.g., Sakurai J J 1994 {\it Modern Quantum Mechanics} (Addison-Wesley Publishing Company)

\bibitem{j}
See, e.g., Joachain C J 1975 {\it Quantum Collision Theory} (North-Holland Publishing Company)


\end{thebibliography}
\end{document}